\documentclass[a4paper,12pt]{article}
\usepackage{amsmath, amssymb, amsthm}
\usepackage{fullpage}
\usepackage{hyperref}
\usepackage{graphicx}
\usepackage{parskip}
\usepackage{color}
\usepackage{appendix}
\usepackage[percent]{overpic}

\numberwithin{equation}{section}

\begin{document}
\title{Skyrmion and Baby Skyrmion Formation from Domain Walls}
\author{Thomas Winyard\\[10pt]
{\em \normalsize Department of Mathematical Sciences, }\\{\em \normalsize Durham University, Durham, DH1 3LE, U.K.}\\[10pt]
{\normalsize t.s.winyard@durham.ac.uk}}

\maketitle
\vspace{15pt}
\begin{abstract}
We numerically simulate the formation of $(2+1)$-dimensional baby Skyrmions and $(3+1)$-dimensional $SU(2)$ Skyrmions from domain wall collisions. It has been suggested that Skyrmion, anti-Skyrmion pairs can be produced from the interaction of two domain walls. This is confirmed, however it is also demonstrated that the process can require quite precise conditions. An alternative, more stable, formation process is proposed as the interaction of more than two segments of domain wall. This is simulated, requiring far less constraints on the initial conditions used. 

Finally domain wall networks are considered, demonstrating how Skyrmions may be produced in a complex dynamical system. We show that the local topological charge configurations, formed within the system, are countered by opposite winding on the boundary of the system to conserve topological charge.
\end{abstract}

\pagebreak

\section{Introduction}
The Skyrme model \cite{Skyrme:1961vq} is a (3+1)-dimensional theory that admits soliton solutions, called Skyrmions, which represent baryons. This has been well studied \cite{bible} with solutions calculated for a large range of charges \cite{Battye:2001qn}.

The baby (or planar) Skyrme model \cite{Piette:1994ug} is the (2+1)-dimensional analogue of the Skyrme model. Baby Skyrmions are also manifest in their own right in condensed matter physics, such as in ferromagnetic quantum Hall systems \cite{Sondhi:1993zz}, and more recently observed in chiral ferromagnets \cite{Yu}. 

In this paper we consider both the $(2+1)$ baby Skyrme model and the full $(3+1)$ $SU(2)$ Skyrme model. We simulate the collisions of domain walls in such a way as to form stable soliton anti-soliton pairs. Normally domain walls will annihilate, however if they interact in such a way as to produce the correct winding in the target space, then soliton anti-soliton pairs can be formed. In the $(2+1)$ model, this consists of the domain walls intersecting to form a ring, with the phase changing by some multiple of $2\pi$ around the ring. In the $(3+1)$ full Skyrme model, there is an additional field and dimension which is needed to wind correctly, this time with the domain walls forming a spherical object in the physical space.

There is a large amount of increased interest in how solitons can be formed, especially in the baby Skyrme model, due to it's proposal for use in spintronics and condensed matter memory systems \cite{sampaio2013nucleation,iwasaki2013current}. There is also interest in the interaction of large domain wall systems and Skyrmions, which is discussed later.

\section{(2+1) Baby Skyrme Model}
The baby Skyrme model has the form of a non-linear modified sigma model, described by the Lagrangian density
\begin{equation}
\mathcal{L}=\frac{1}{2}\partial_\mu \boldsymbol{\phi}\cdot\partial^\mu \boldsymbol{\phi}-\frac{\kappa^2}{4}\left(\partial_\mu\boldsymbol{\phi}\times\partial_\nu\boldsymbol{\phi}\right)\cdot \left(\partial^\mu \boldsymbol{\phi} \times\partial^\nu\boldsymbol{\phi} \right)-m^2\left(1-\phi_3^2\right),
\label{lagrange}
\end{equation}
where greek indicies run over time and spatial dimensions ($\mu = 0,1,2$) and $\boldsymbol{\phi}(\mathbf{x},t)$ is a unit vector field, $\boldsymbol{\phi}=(\phi_1,\phi_2,\phi_3)$. For finite energy we require $\boldsymbol{\phi}$ to be a vacuum at spatial infinity, hence it can be viewed as a map from the compactified physical space, $\mathbb{R}^2 \cup \{\infty\}=\textit{S}\,^2$, to the target space $\textit{S}\,^2$. Since the second homotopy group $\pi_2(\textit{S}\,^2)=\mathbb{Z}$ there is an integer valued winding number associated to the map, which is characterised by the topological charge,
\begin{equation}
B=-\frac{1}{4\pi}\int \boldsymbol{\phi}\cdot(\partial_1 \boldsymbol{\phi}\times \partial_2 \boldsymbol{\phi})\,\mathrm{d}^2\mathbf{x}.
\end{equation}

Due to our choice of potential there are two choices of vacuum denoted $\boldsymbol{\phi}_\pm$,

\begin{equation}
\boldsymbol{\phi}_\infty = \lim_{\left|\boldsymbol{x}\right|\rightarrow\infty}\boldsymbol{\phi}\left(\boldsymbol{x},t\right) = \boldsymbol{\phi}_\pm = \left(0,0,\pm 1\right).
\end{equation}

The inclusion of the mass term breaks the $O(3)$ symmetry to $O(2)\times Z_2$, the selection of a vacuum on the boundary of the physical space then breaks the symmetry further to an $O\left(2\right)$ symmetry. While other mass terms exist, there needs to be at least two choices of distinct discontinuous vacua to allow domain walls to form. In fact for the purposes of baby Skyrmion formation, it is optimal for these disconnected vacuua to lie on antipodal points of the target $S^2$ space.

Domain walls are walls of energy that interpolate from one vacuum to another (here $\boldsymbol{\phi}_\pm$). Normally they have no topological charge in and of themselves, however special domain wall solutions have been found that do contain winding \cite{Nitta:2012xq,Kobayashi:2013ju}. This winding is given stability due to the constraining domain wall.

\begin{figure}
\begin{center}
\begin{tabular}{c c c}
\includegraphics[scale=0.09,natwidth=1000,natheight=1000]{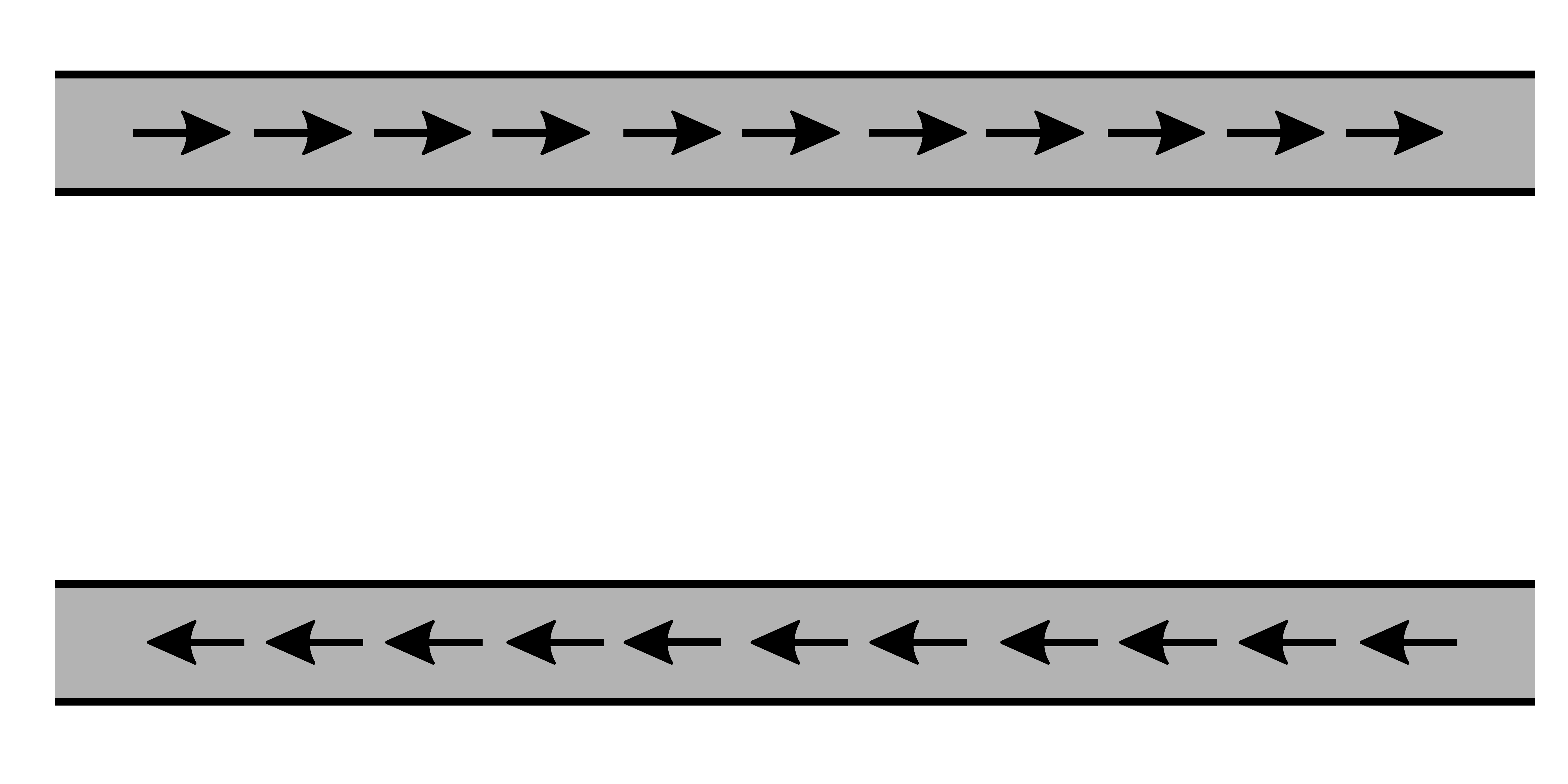} & \includegraphics[scale=0.09,natwidth=1000,natheight=1000]{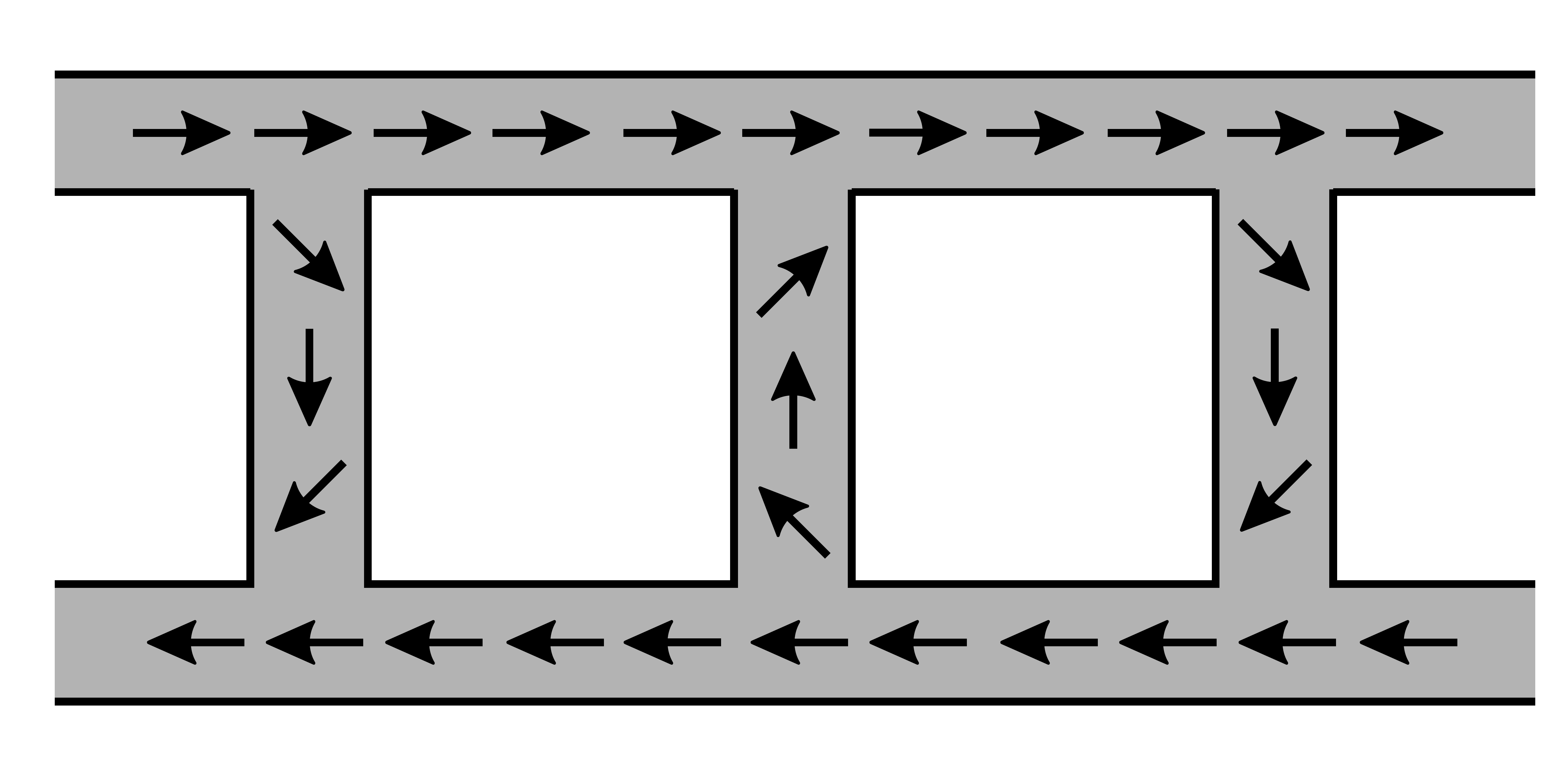} &
\includegraphics[scale=0.09,natwidth=1000,natheight=1000]{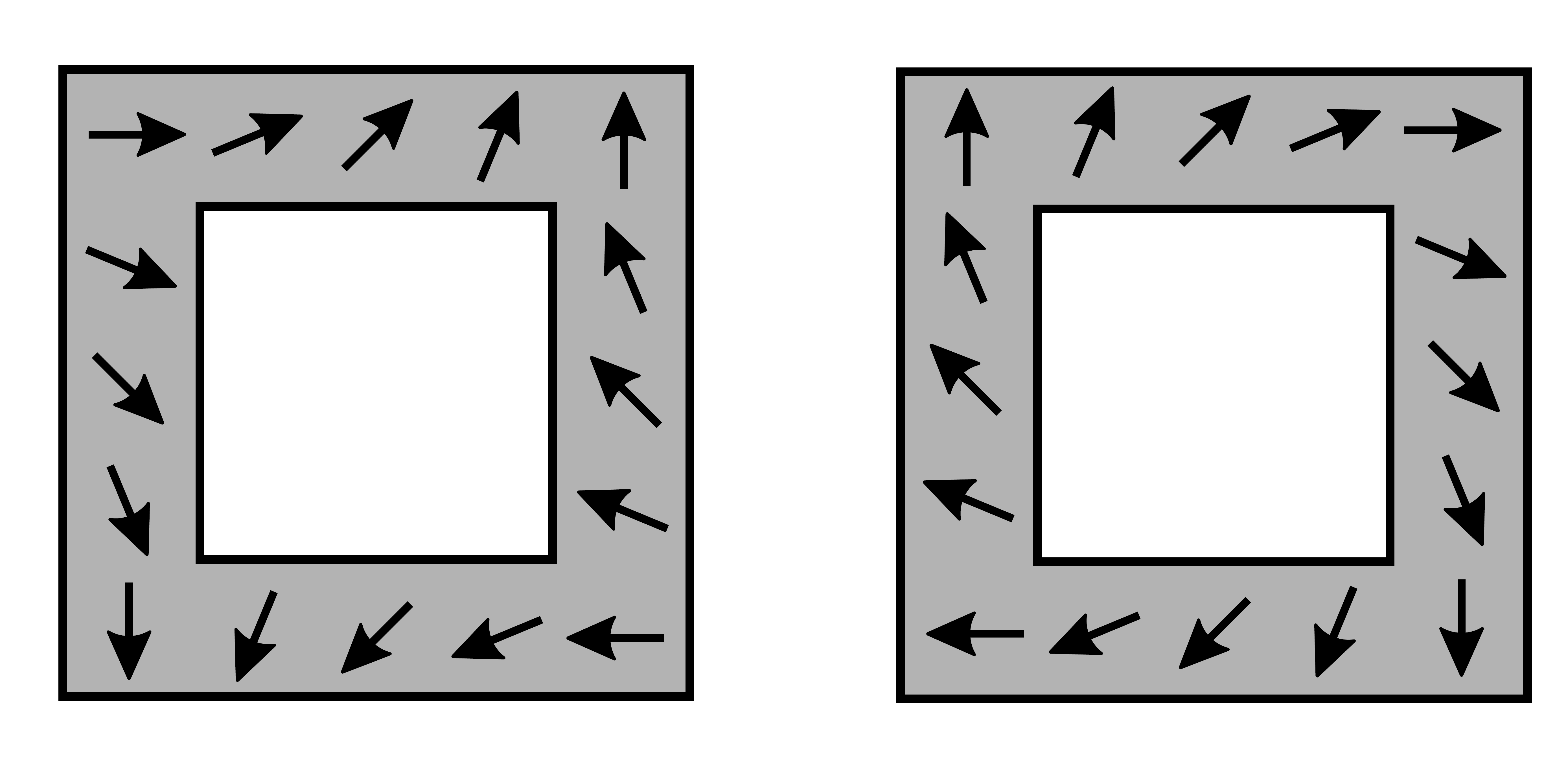}
\end{tabular}
\end{center}
\caption{Annihilation of two domain walls by the formation of bridges that interpolate the phase of the walls, forming in such a way as to produce a winding effect. With the correct winding, a soliton anti-soliton pair are formed}
\label{PlanarFormation}
\end{figure}

It has been suggested that baby Skyrmion solutions can be formed from domain wall collisions \cite{PhysRevA.85.053639}. If the domain walls collide in such a way as to form channels between them, with the correct winding round the loops formed, then baby Skyrmion anti-Skyrmion pairs can be formed, as shown in figure \ref{PlanarFormation}. Note that this doesn't break topological charge invariance as a soliton anti-soliton pair has been formed. 

If we consider the process in terms of the target space, the domain walls traverse between the two antipodal vacua of the target $S^2$. The domain walls intersect at points along their length. To achieve this, a bridge must form that sweeps around the target sphere to match the field configuration of the opposite domain wall. The bridge essentially has a choice, it can sweep one of two ways around the target space. If you have two bridges form adjacent, that wind round the target space in the opposite direction, then when they coalesce, a loop has formed that winds round the target space once.

While this has formed a baby Skyrmion or anti-Skyrmion (depending on which way round the bridges formed), topological charge invariance wont be broken. This is due to domain walls being infinitely long or forming in loops, which segment areas of space into different vacuum. If we return to the example presented before where the domain walls meet and form only two bridges, which wind correctly. We can consider these bridges as physical objects that propagate along the walls in both directions. If the walls are infinitely long then the bridges will meet opposite bridges on both sides forming a soliton anti-soliton pair. Hence a chain of soliton anti-soliton pairs can form. If the domain walls are loops, then the bridges will meet at the initial interaction point, however they will then propagate around the walls and meet again to produce the opposite winding.

\section{Baby Skyrmion Formation Examples}
Simulations of the nonlinear time-dependent PDE that follows from the variation of \ref{lagrange} were performed using a fourth order Runge-Kutta method, on a grid of 501x501 grid points, with $4^{th}$ order finite difference derivatives. We used Neumann boundary conditions (the spatial derivative normal to the boundary vanishes), which allows the domain walls to move unhindered. In theory the domain walls are infinitely long (or formed from systems of domain wall loops), however in any simulation or experimental system we deal with a finite segment. 

\begin{figure}
\begin{center}
\begin{tabular}{c c c c}
\includegraphics[scale=0.35,natwidth=1000,natheight=1000]{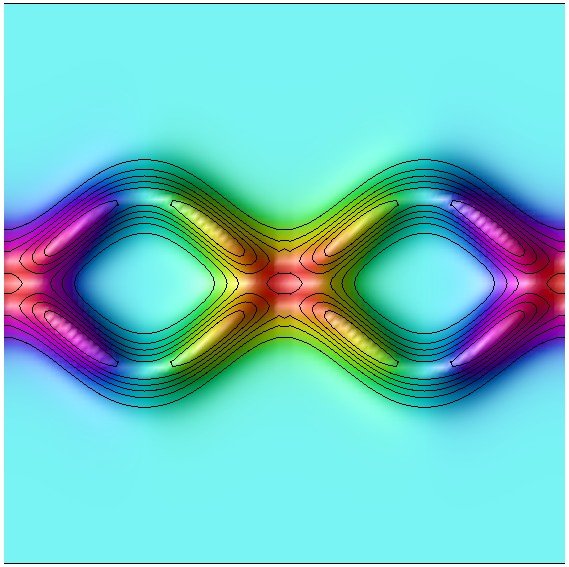} & \includegraphics[scale=0.35,natwidth=1000,natheight=1000]{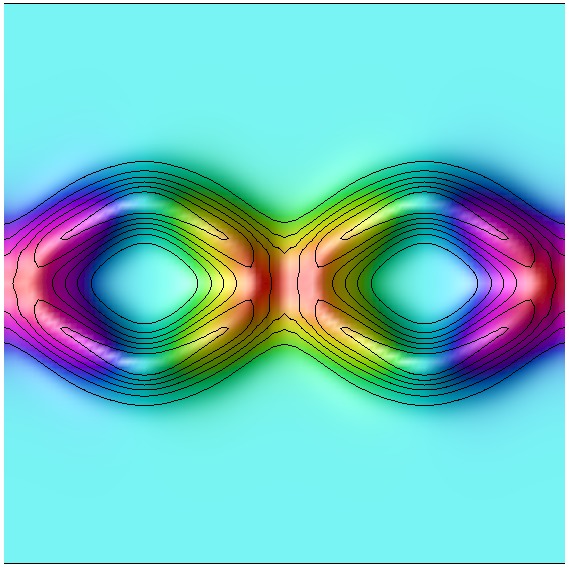} & \includegraphics[scale=0.35,natwidth=1000,natheight=1000]{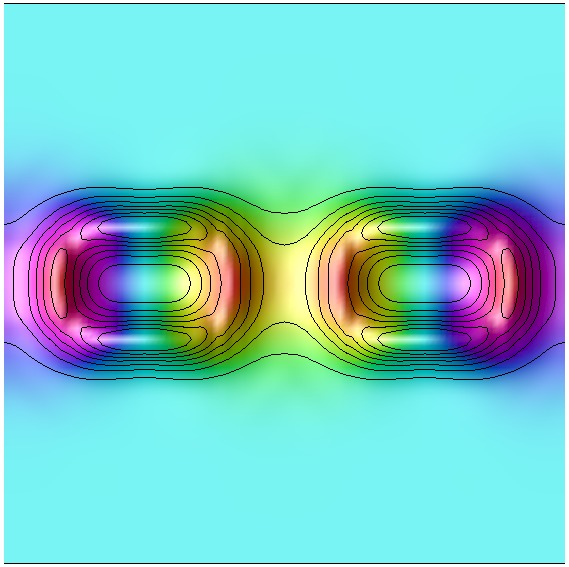} &
\includegraphics[scale=0.35,natwidth=1000,natheight=1000]{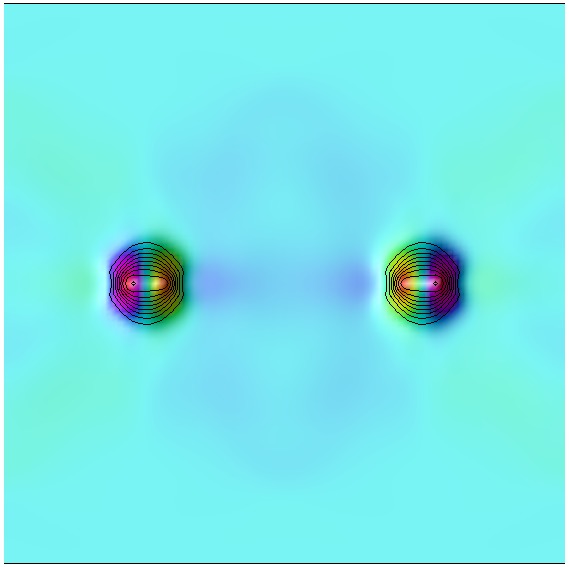}\\
t = 0.4 & t = 1.6 & t = 4.4 & t=10.4 \\
\includegraphics[scale=0.35,natwidth=1000,natheight=1000]{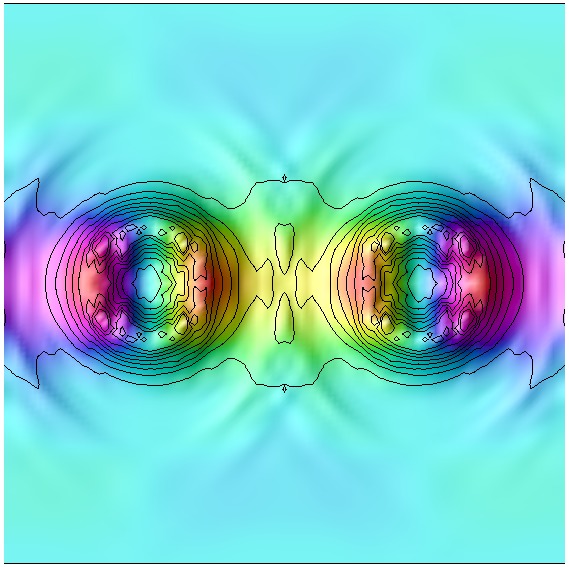} & \includegraphics[scale=0.35,natwidth=1000,natheight=1000]{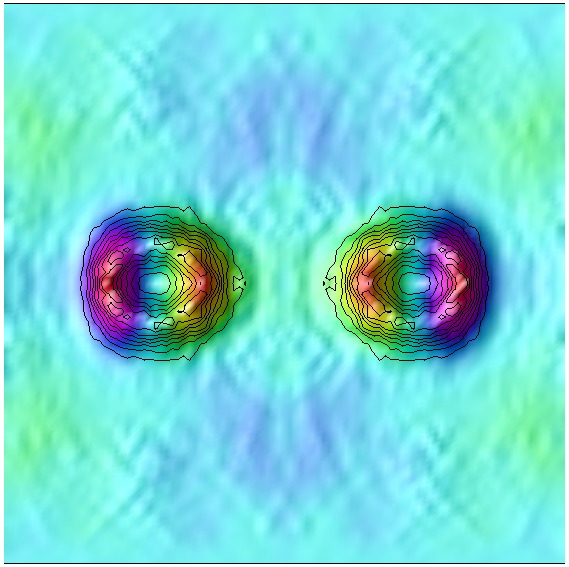} & \includegraphics[scale=0.35,natwidth=1000,natheight=1000]{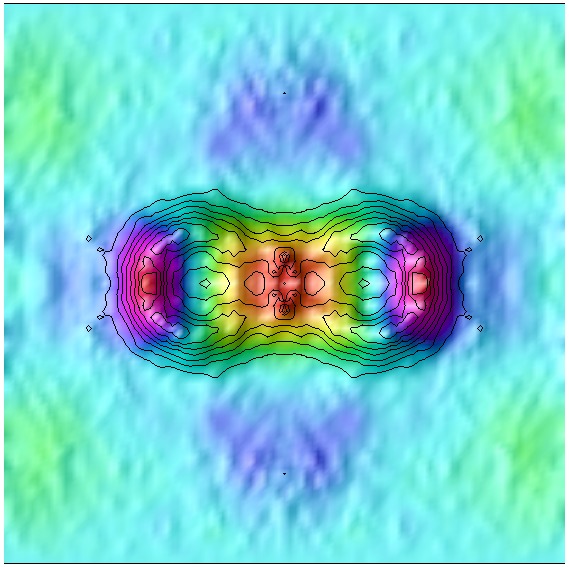} &
\includegraphics[scale=0.35,natwidth=1000,natheight=1000]{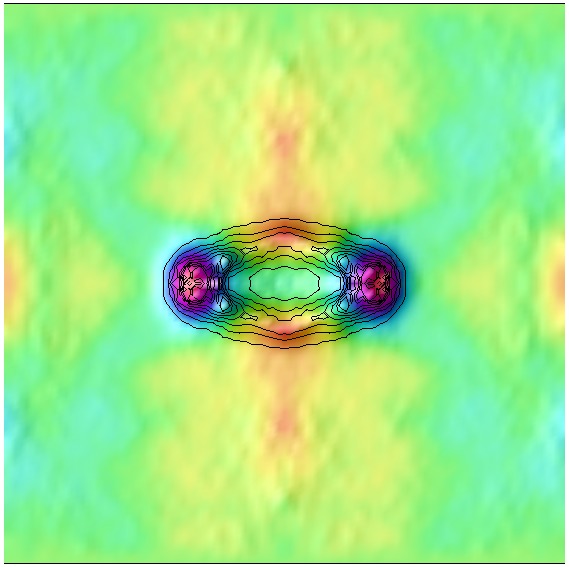}\\
t = 19.6 & t = 42.4 & t = 76.4 & t = 81.2
\end{tabular}
\caption{An energy density contour plot of 2 domain walls that have been perturbed to simulate the forming of bridges. The bridges are oriented to cause the fields to wind correctly to form a soliton anti-soliton pair. The two solitons initially reduce in size then they attract and annihilate. Due to the large quantities of energy involved the solitons oscillate in size while attracting until they ultimately annihilate. The plot is coloured by the value of the $\phi_1$ field.}
\label{PlanarFormation1}
\end{center}
\end{figure}

We first simulate the process outlined in figure \ref{PlanarFormation}, however this requires our initial conditions to be highly constrained. Domain walls by their nature want to minimise their length (become straight in $\mathbb{R}^2$). They also want to match the phase of any other incident walls. You can see the production process of a soliton anti-soliton pair from two domain walls in figure \ref{PlanarFormation1}. If the phases are correctly wound, then the pair forms and ultimately annihilates. In this simulation we have perturbed two standard straight domain walls to simulate bridges forming and winding the phases round. However this doesn't naturally happen in a simulation, as the walls will normally collide across their length, having equalised their phases across the length of the walls. It is possible that this could occur in a domain wall network, where there are more interactions occurring with other domain walls in the system. It is likely the walls could then meet on a scale far larger than the size of a Skyrmion and hence the bridges formed would not affect each other initially, allowing opposite directions around the target space to be selected.

\begin{figure}
\begin{center}
\begin{tabular}{c c c c c}
\includegraphics[scale=0.25,natwidth=1000,natheight=1000]{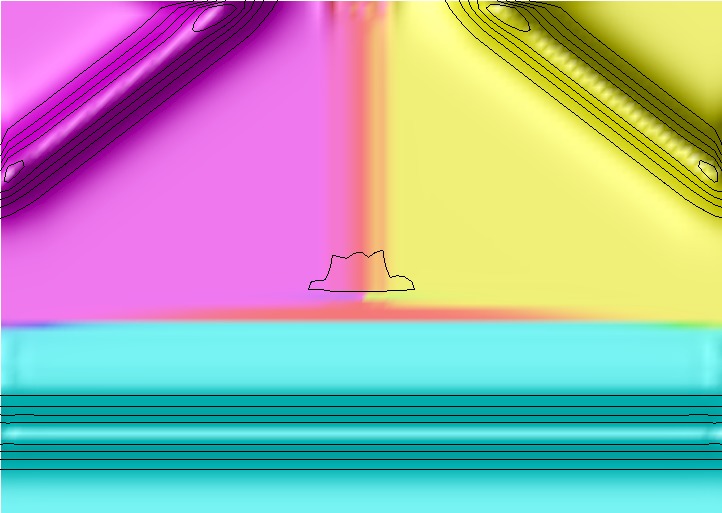} & \includegraphics[scale=0.25,natwidth=1000,natheight=1000]{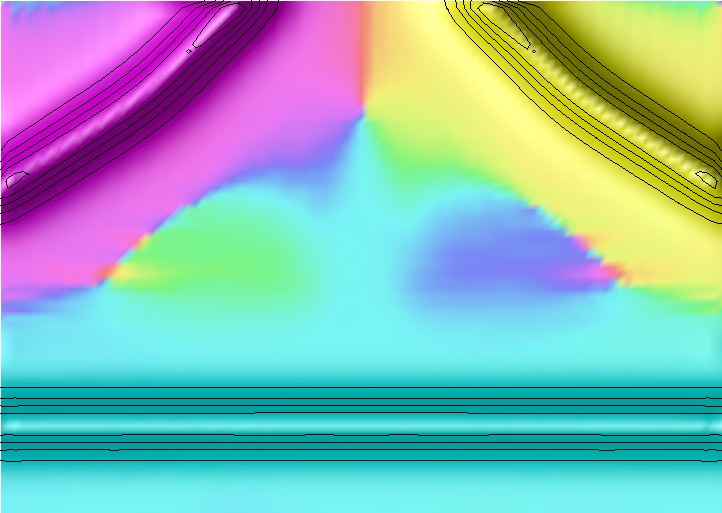} & \includegraphics[scale=0.25,natwidth=1000,natheight=1000]{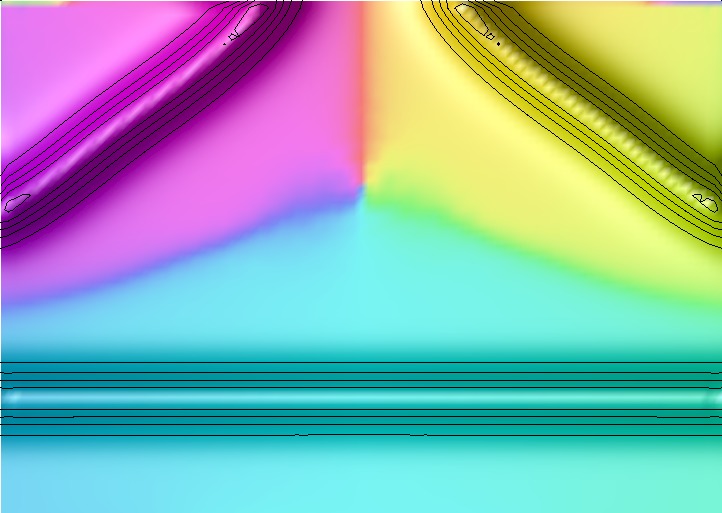} &
\includegraphics[scale=0.25,natwidth=1000,natheight=1000]{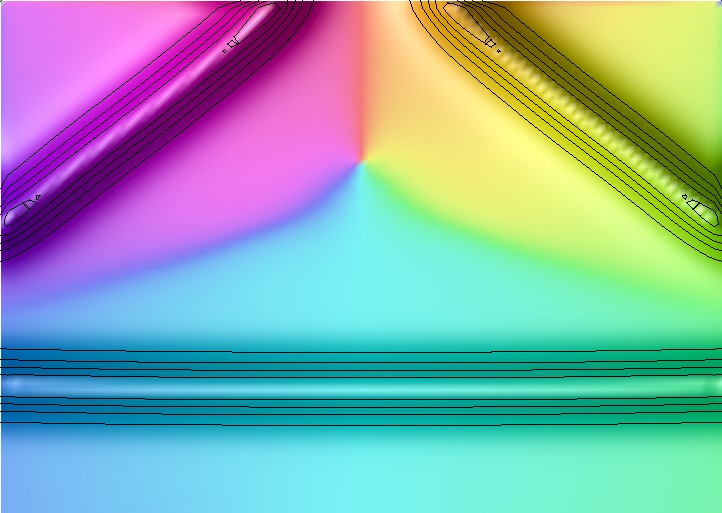} & \includegraphics[scale=0.25,natwidth=1000,natheight=1000]{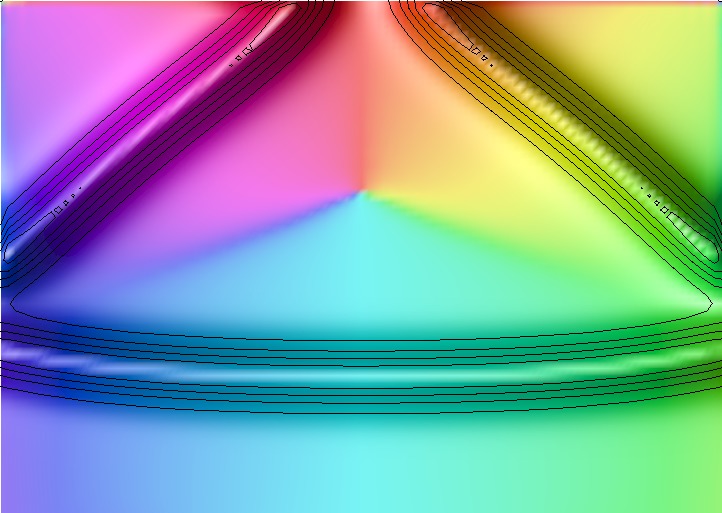}\\
t = 0 & t = 25 & t = 100 & t = 150 & t = 200 \\
\includegraphics[scale=0.25,natwidth=1000,natheight=1000]{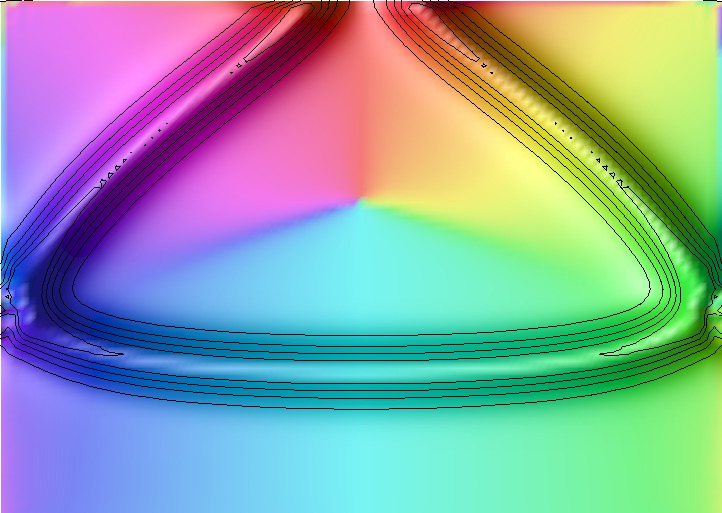} & \includegraphics[scale=0.25,natwidth=1000,natheight=1000]{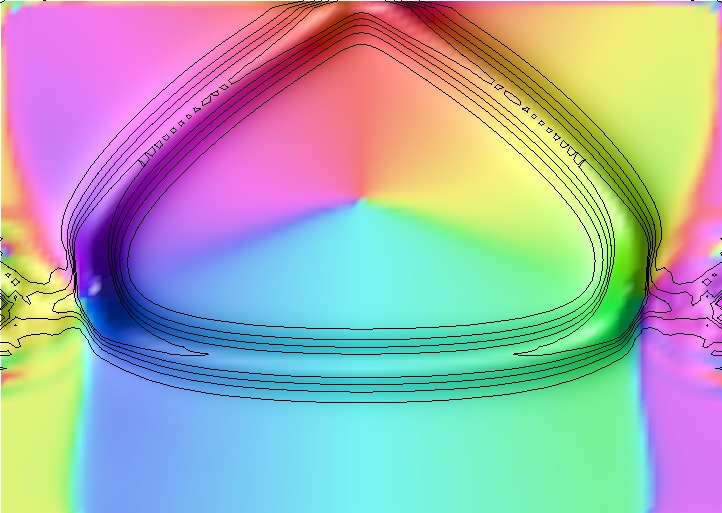} & \includegraphics[scale=0.25,natwidth=1000,natheight=1000]{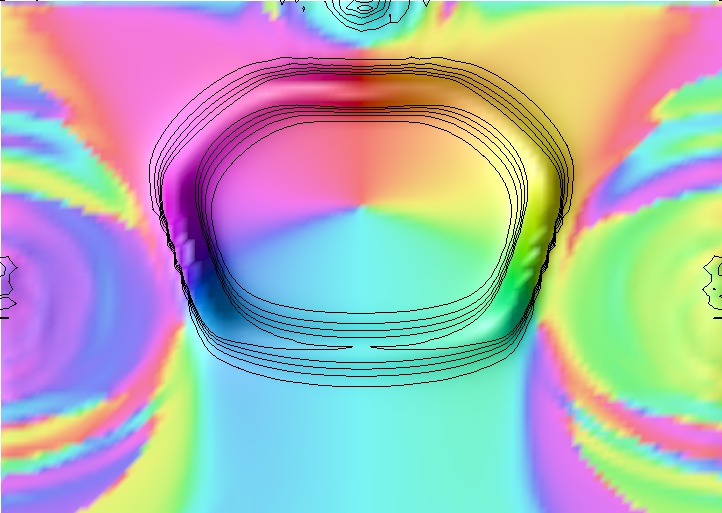} &
\includegraphics[scale=0.25,natwidth=1000,natheight=1000]{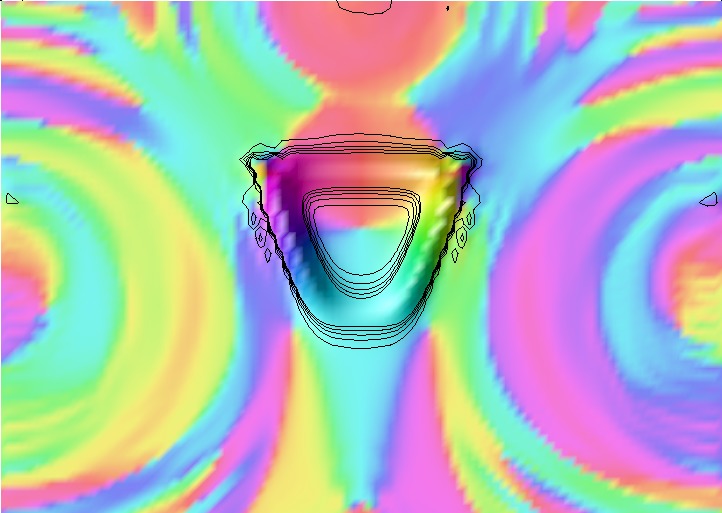} & \includegraphics[scale=0.25,natwidth=1000,natheight=1000]{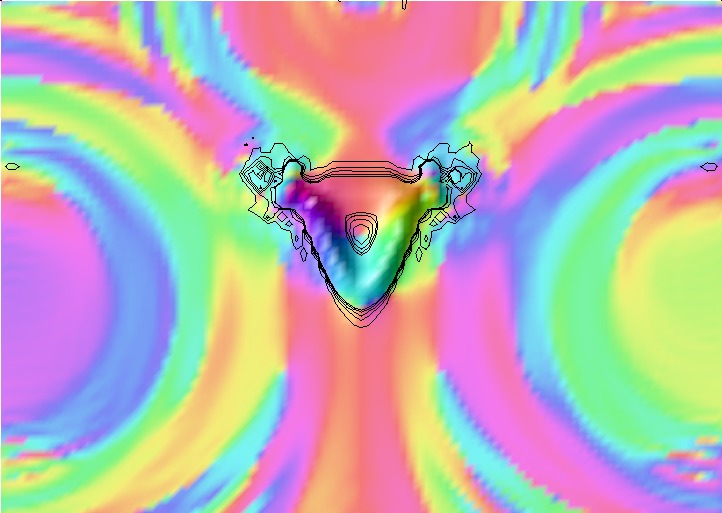}\\
t = 205 & t = 210 & t = 217.5 & t = 222.5 & t = 229.5
\end{tabular}
\caption{Energy density plot for three incident domain walls with different phases. The walls attract, attempting to equalise their phases on both sides. This leads to the correct winding for a soliton, once the walls have interacted. The plot is coloured by the phase $\theta = \tan^{-1}\frac{\phi_2}{\phi_1}$.}
\label{3walls}
\end{center}
\end{figure}

\begin{figure}
\begin{center}
\begin{tabular}{c c c c c}
\includegraphics[scale=0.25,natwidth=1000,natheight=1000]{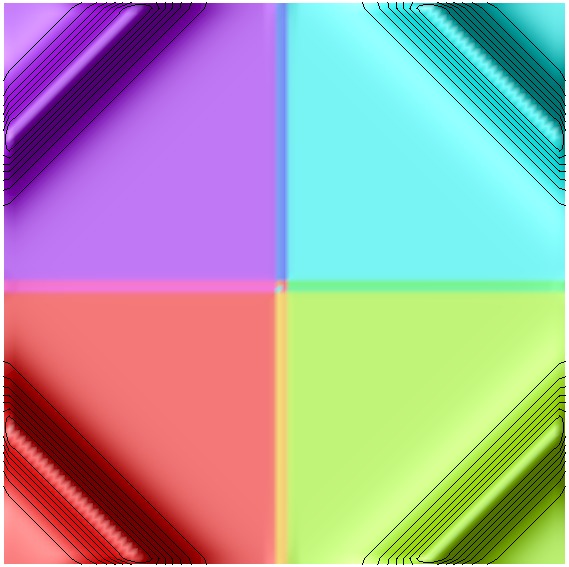} & \includegraphics[scale=0.25,natwidth=1000,natheight=1000]{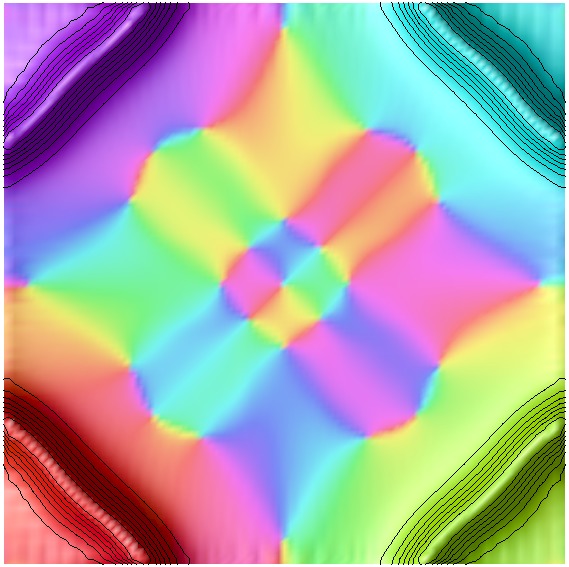} & \includegraphics[scale=0.25,natwidth=1000,natheight=1000]{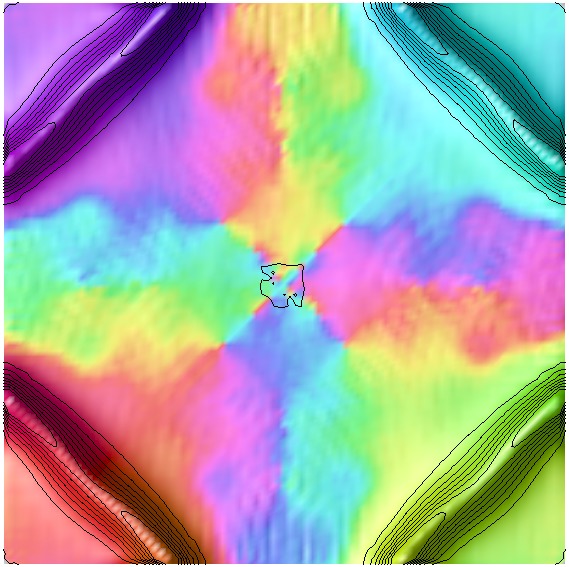} &
\includegraphics[scale=0.25,natwidth=1000,natheight=1000]{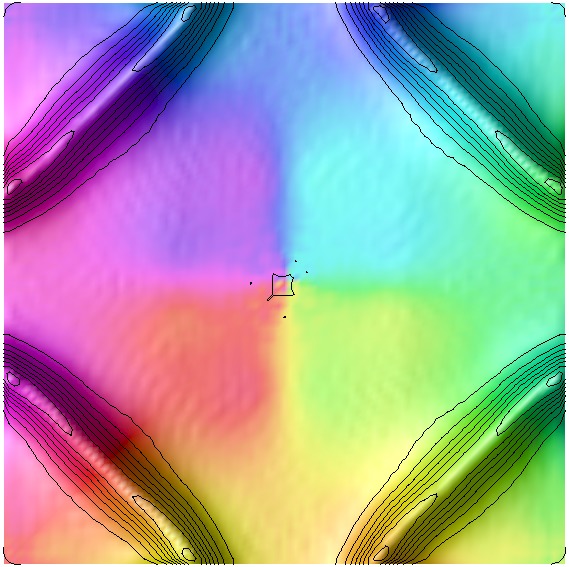} & \includegraphics[scale=0.25,natwidth=1000,natheight=1000]{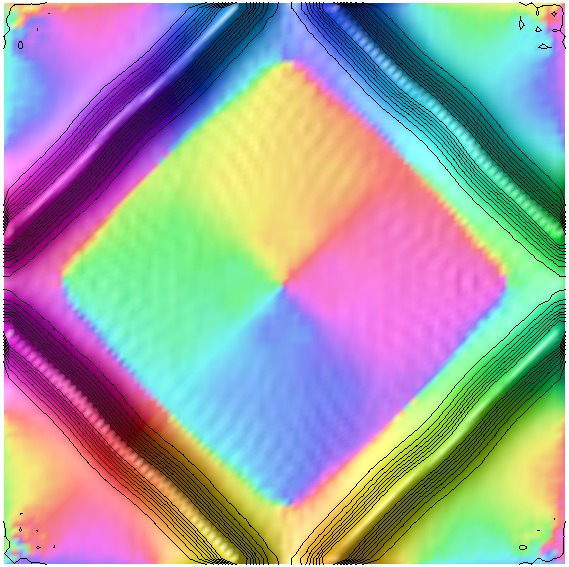}\\
t = 0 & t = 100 & t = 200 & t = 250 & t = 266 \\
\includegraphics[scale=0.25,natwidth=1000,natheight=1000]{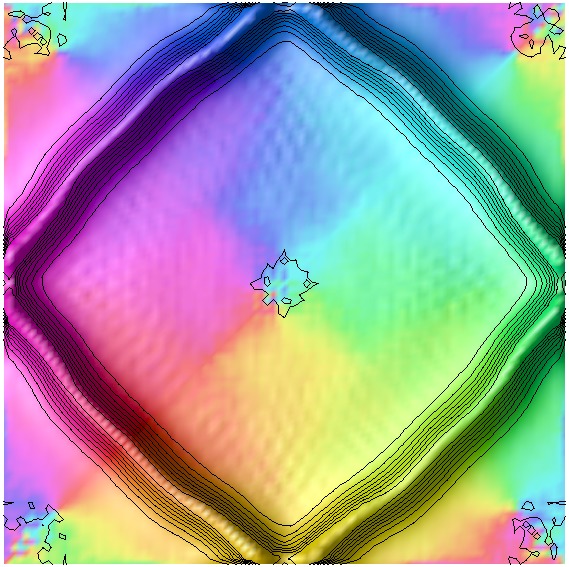} & \includegraphics[scale=0.25,natwidth=1000,natheight=1000]{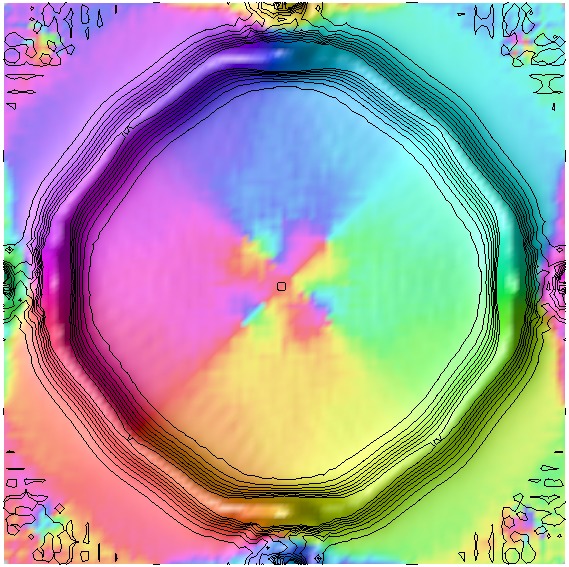} & \includegraphics[scale=0.25,natwidth=1000,natheight=1000]{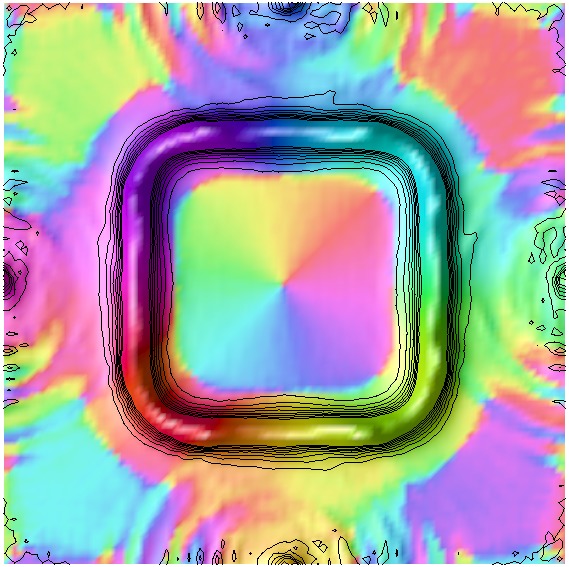} &
\includegraphics[scale=0.25,natwidth=1000,natheight=1000]{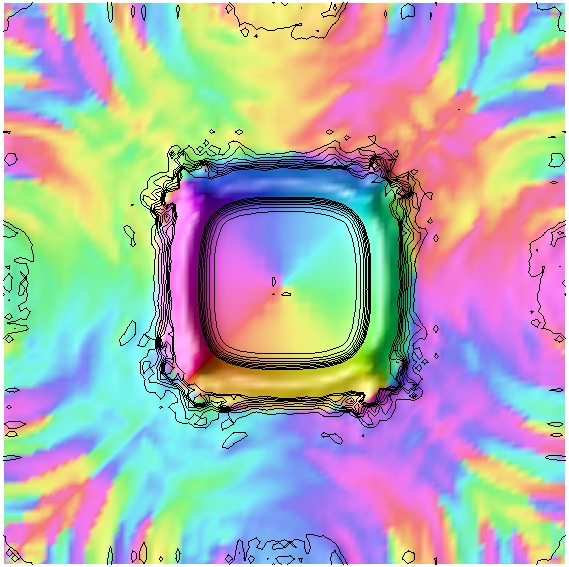} & \includegraphics[scale=0.25,natwidth=1000,natheight=1000]{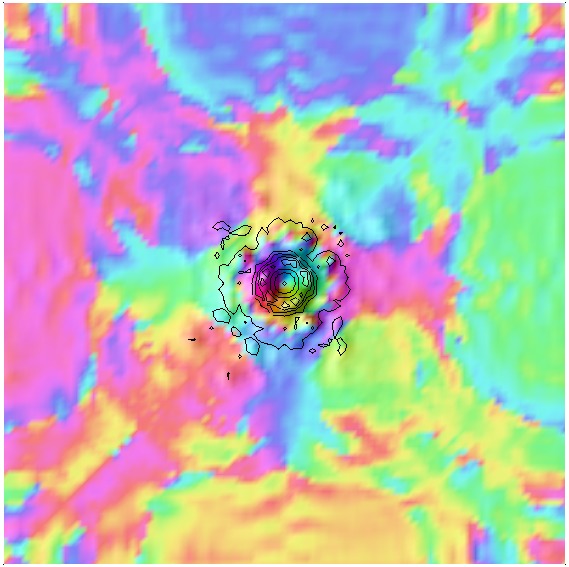}\\
t = 270 & t = 274 & t = 282 & t = 288 & t = 296
\end{tabular}
\caption{Energy density plot of four incident domain walls with different phases. They match phases and create the correct winding. This could create a charge 2 solution if two adjacent wall phases were swapped, due to the field needing to wind twice. It is likely a larger scale is needed for this to occur however. The plot is coloured by the phase $\theta = \tan^{-1}\frac{\phi_2}{\phi_1}$.}
\label{4walls}
\end{center}
\end{figure}

To produce a production process in which we don't have to heavily constrain the initial conditions, we have to add an additional domain wall. The formation process for a single soliton can be seen for 3 domain walls in figure \ref{3walls} and for 4 domain walls in figure \ref{4walls}. Note that we are now only considering the part that forms the soliton, not the matching anti-soliton that should be formed further down the domain wall interaction. 

We observe that the domain walls will try to match their phase with the domain walls on either side of them, causing the phase to partially wind along the length of the wall. Should the phases of each incident domain wall be well separated, then the winding around all the domain walls will produce a single charge soliton, as the walls annihilate with each other.

\begin{figure}
\begin{center}
\begin{tabular}{c c c c c}
\includegraphics[scale=0.25,natwidth=1000,natheight=1000]{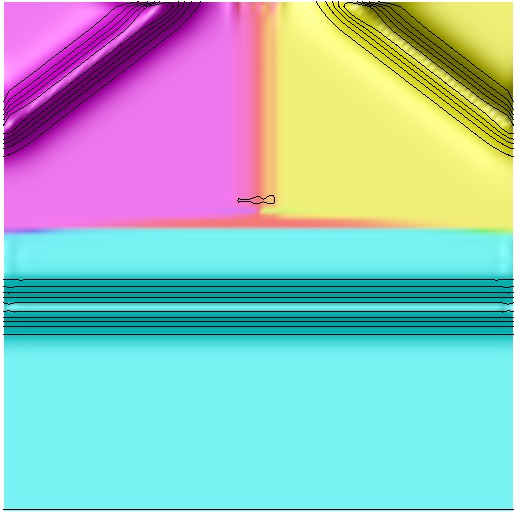} & \includegraphics[scale=0.25,natwidth=1000,natheight=1000]{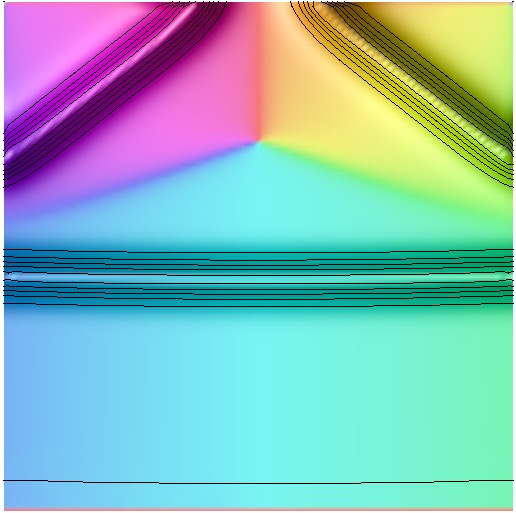} & \includegraphics[scale=0.25,natwidth=1000,natheight=1000]{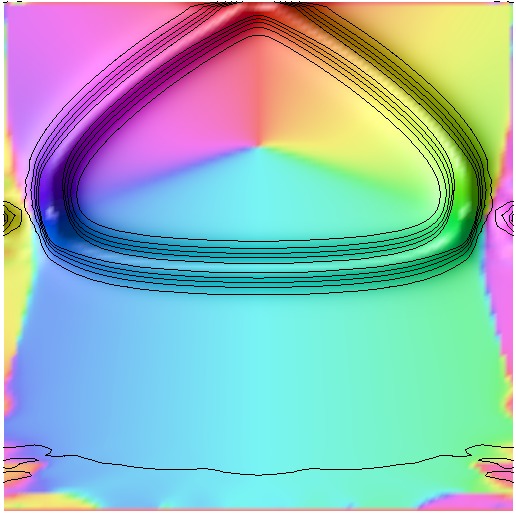} &
\includegraphics[scale=0.25,natwidth=1000,natheight=1000]{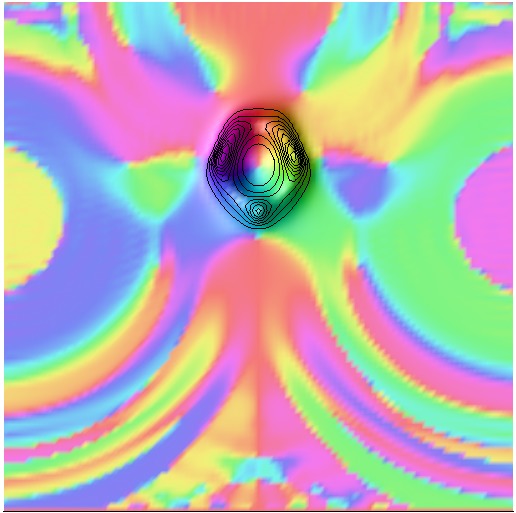} \\
t = 0 & t = 346 & t = 384 & t = 402 \\
\includegraphics[scale=0.25,natwidth=1000,natheight=1000]{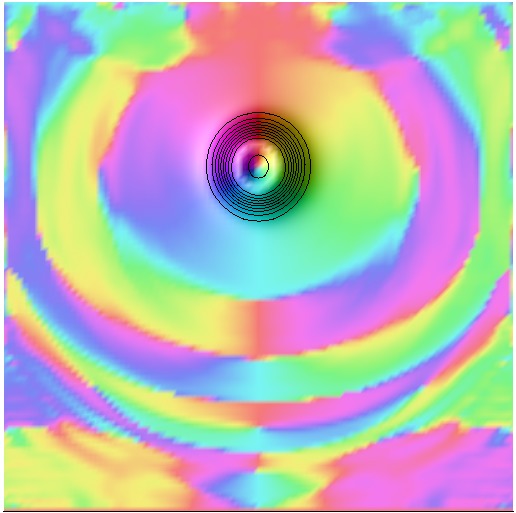} & \includegraphics[scale=0.25,natwidth=1000,natheight=1000]{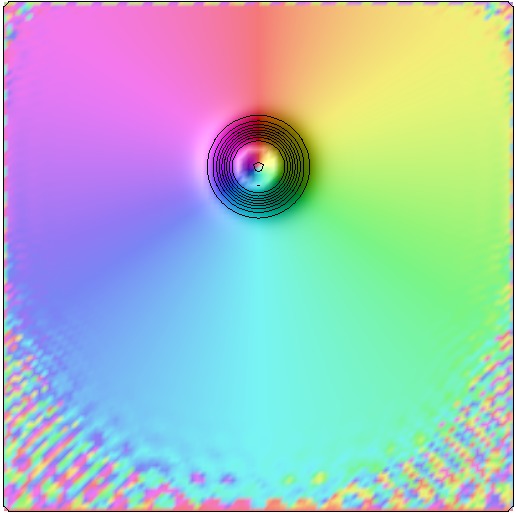} &
\includegraphics[scale=0.25,natwidth=1000,natheight=1000]{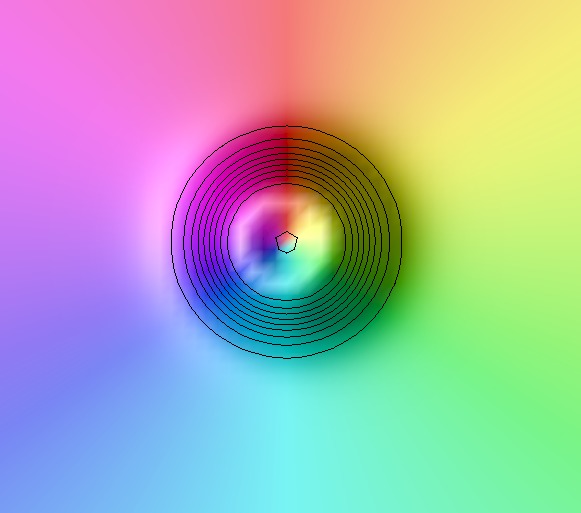} & 
\includegraphics[trim=85 0 75 0,clip,scale=0.4]{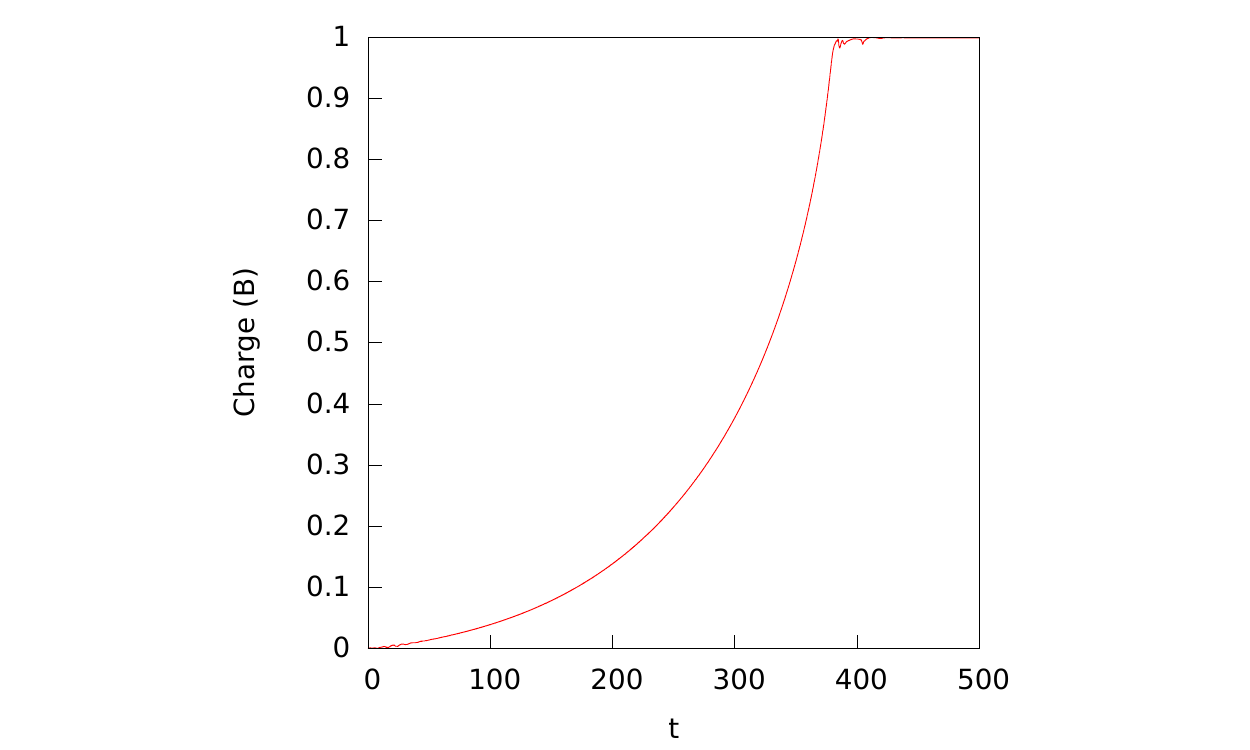} \\
t = 428 & t = 566 & t = 1000 & charge $B$
\end{tabular}
\caption{Energy density plot of three incident domain walls with different phases and heavy damping. They match phases and create the correct winding. The penultimate panel shows a blown up image of the resulting baby Skyrmion and the final panel is the changing topological charge over time.}
\label{3walls_damp}
\end{center}
\end{figure}

The large amount of kinetic energy makes keeping the resulting soliton stable quite challenging, hence the process was repeated with high damping, resulting in figure \ref{3walls_damp}. Here the resulting soliton remains constant and the topological charge has also been plotted showing an increase from $B=0$ to $B=1$. You can see that this has occurred due to a discontinuous deformation made to the system, moving the domain wall away from the boundary of the space. This requires damping to counteract this but allows the topological charge to be artificially changed. 

In figure \ref{4walls} you can observe a single soliton being formed by 4 domain walls in a similar manner. If you wind the incident phases in the 4 wall case round the target space twice it should be possible to form a charge 2 soliton instead of a charge 1. However you will require the same stringent initial conditions as with the 2 domain walls forming a single soliton in figure \ref{PlanarFormation1}. Alternatively a large scale would be required to ensure the bridges don't interact before they have chosen a route round the target space. However if you consider the meeting of 5 domain walls, then the winding can be easily created for a charge 2 soliton. This result should continue for higher numbers of incident domain walls, assuming the phases are distributed in the correct manner.

This leads us to conclude that multiple interacting domain walls have a higher chance to produce a baby Skyrmion, rather than the highly constrained requirements of two domain walls annihilating. This idea can be put into practice in a condensed matter system. Here the formation of baby Skyrmions at will is of great interest. If three domain walls were to meet at a bifurcation point (Y-junction) in a system, then the chances of producing a stable soliton would be quite high. The difficulty would arise in having phases that are well separated. This can be achieved by considering a theory that promotes certain phases for domain walls. To achieve this a mass term can be used that breaks the traditional $O(2)$ symmetry of the theory to some dihedral group. Such a potential has been studied in \cite{Jaykka:2011ic,Winyard:2013ada} (shown in \ref{D_Npot}) that breaks the symmetry to a $D_N$ subgroup, a sensible choice would be to set $N$ equal to the number of incident walls, ensuring their phases are well separated. 

\begin{equation}
V[\boldsymbol{\phi}]=\left|1-(\phi_1+i\phi_2)^N\right|^2(1-\phi_3).
\label{D_Npot}
\end{equation}

This term has the correct $D_N$ symmetry however there is a subtlety in that it creates new possible vacua, that are no longer antipodal. Hence domain walls can now form at the $N$ points on the equator of the target space where $(\phi_1+i\phi_2)^N = 1$.

For the optimal potential we must remove these vacua while keeping the dihedral symmetry and add two antipodal vacua, which is quite simple mathematically,

\begin{equation}
V[\boldsymbol{\phi}]=\left|\alpha-(\phi_1+i\phi_2)^N\right|^2(1-\phi_3^2)
\end{equation}

where $\alpha \gg 1$. In practice this potential seems somewhat artificial, though it serves the purpose of demonstrating how a system should be constrained, to allow increased production of the correct winding to form baby Skyrmions.

\section{Domain Wall Systems}
One system in which a baby Skyrmion can be formed is a system of interacting domain walls. This consists of vacua separated by domain walls in loops, that want to annihilate to reduce the energy of the system. We will consider simple interactions, that may occur between loops of domain walls, in such a way as to create baby Skyrmions. Note that as we are no longer considering infinite objects, the energy of the system is now finite. It also allows us to set the boundary of our system to be the same value $\phi_+$, and Neumann boundary conditions are no longer required.

\begin{figure}
\begin{center}
\begin{tabular}{c c c}
\begin{overpic}[scale=0.08]{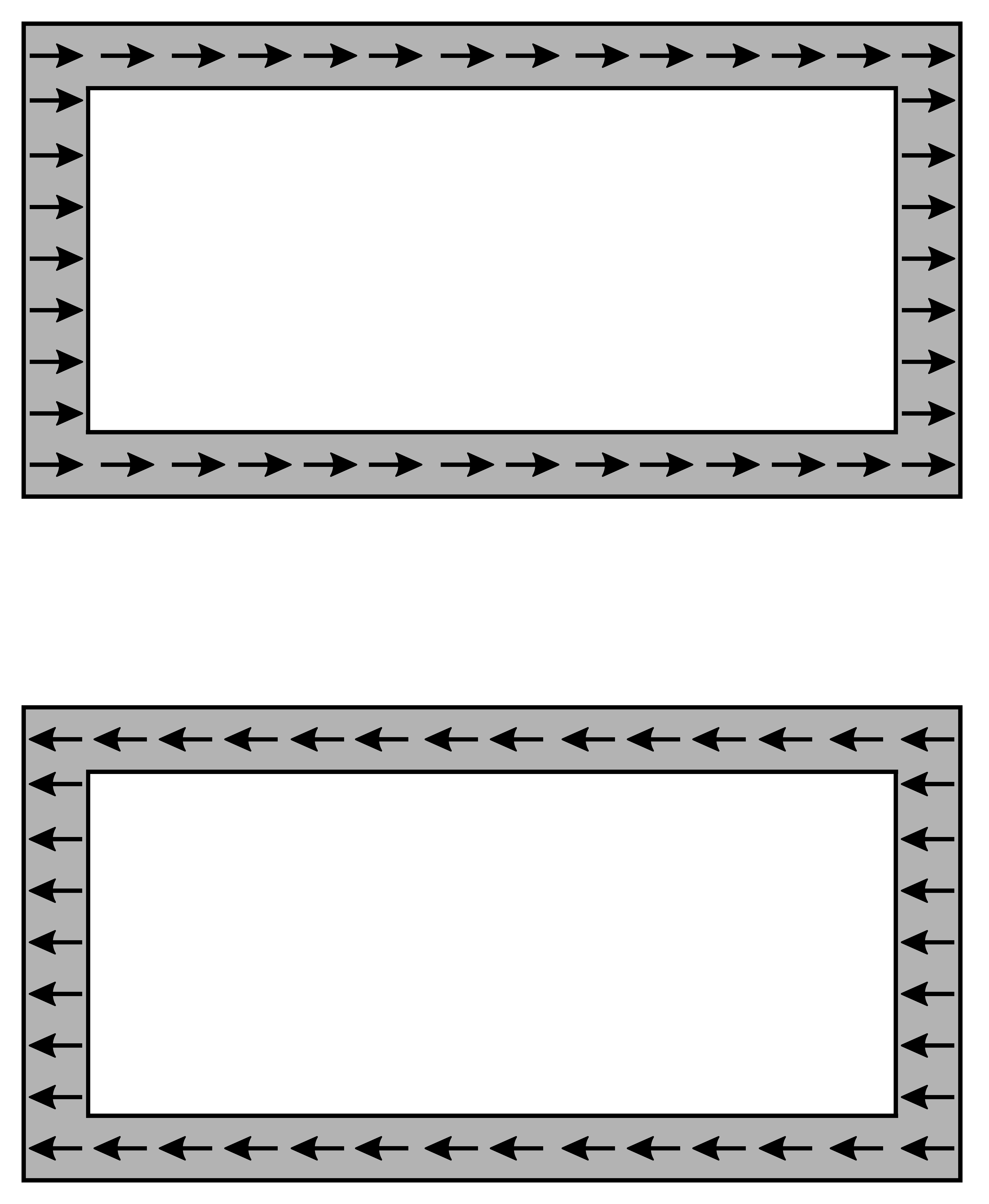}
 \put (38,75) {\large$\displaystyle\phi_-$}
 \put (38,47) {\large$\displaystyle\phi_+$}
 \put (38,18) {\large$\displaystyle\phi_-$}
\end{overpic}
 & \begin{overpic}[scale=0.08]{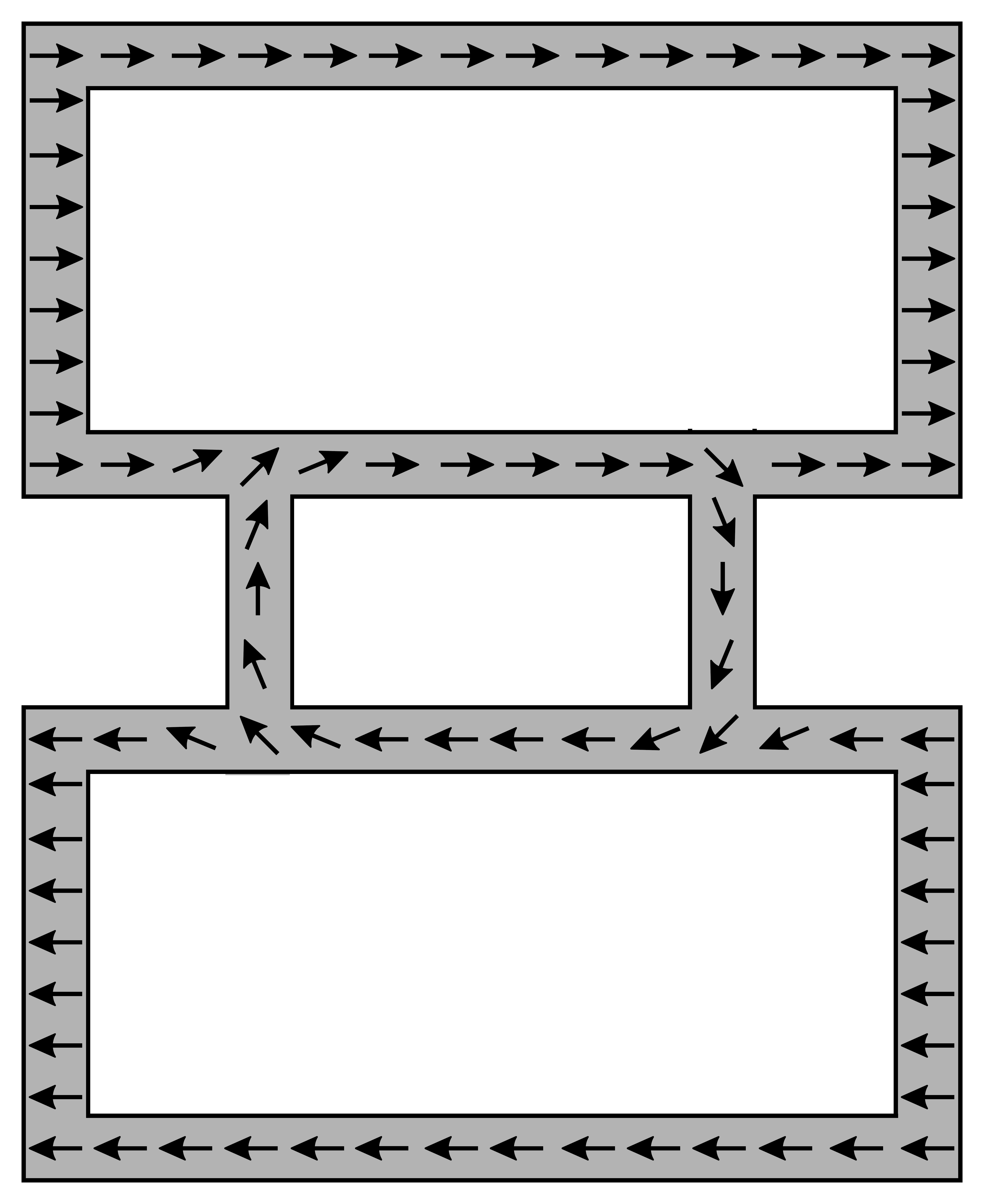}
 \put (38,75) {\large$\displaystyle\phi_-$}
 \put (38,47) {\large$\displaystyle\phi_+$}
 \put (38,18) {\large$\displaystyle\phi_-$}
 \put (68,47) {\large$\displaystyle\phi_+$}
 \put (6,47) {\large$\displaystyle\phi_+$}
\end{overpic} &
\begin{overpic}[scale=0.08]{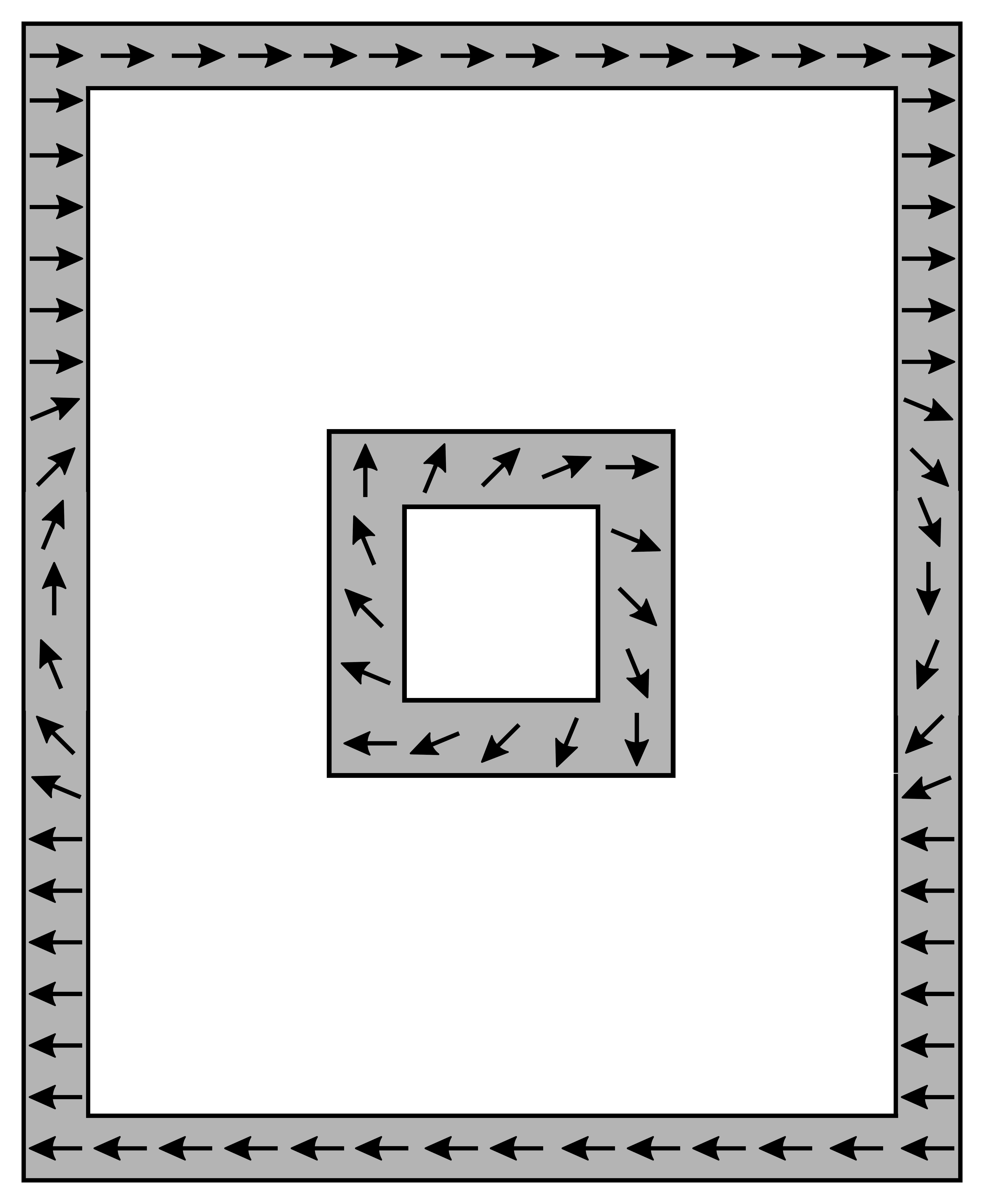}
 \put (38,75) {\large$\displaystyle\phi_-$}
 \put (38,47) {\large$\displaystyle\phi_+$}
 \put (38,18) {\large$\displaystyle\phi_-$}
\end{overpic}
\end{tabular}
\end{center}
\caption{Annihilation of two domain wall bubbles. Bridges form, interpolating between the phase of the two domain walls that wind correctly to form a Skyrmion. As the bridges annihilate a Skyrmion forms and some fractional winding is created on either side of the boundary domain wall. The fractional winding sections on the domain wall cancel the winding of the Skyrmion as the domain wall interpolates $\phi_3$ in the opposite direction to the interior Skyrmion. The various vacuum regions the domain walls interpolate between are denoted $\phi_\pm$.}
\label{PlanarLoopFormation}
\end{figure}

A single interaction has been drawn in figure \ref{PlanarLoopFormation}, showing how two loops, if they form bridges, can produce a temporary local topological charge density. Note that it may seem again that we have broken topological charge invariance, however the produced Skryrmion winding is counteracted by the winding of the domain wall surrounding it, which winds in the opposite direction. This may not be obvious at first, as the phase winds in the same direction for both objects. However, the surrounding domain wall interpolates $\phi_3$ in the opposite direction, hence producing negative winding to the baby Skyrmion in the centre.

\begin{figure}
\begin{center}
\begin{tabular}{c c c c}
\includegraphics[scale=0.35,natwidth=1000,natheight=1000]{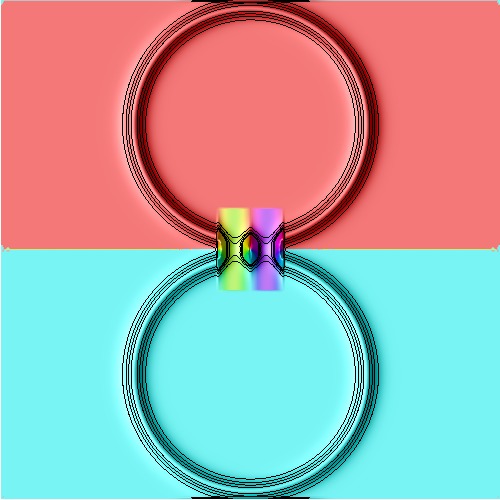} & \includegraphics[scale=0.35,natwidth=1000,natheight=1000]{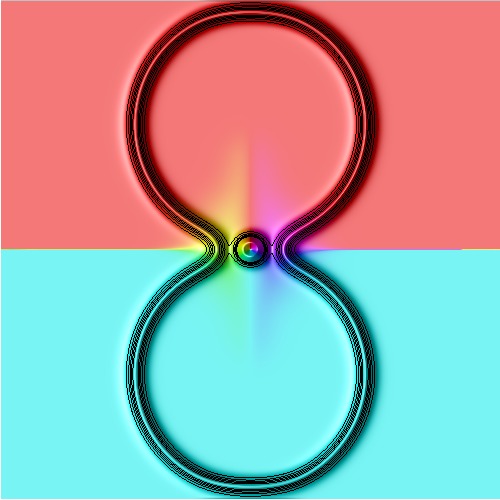} & \includegraphics[scale=0.35,natwidth=1000,natheight=1000]{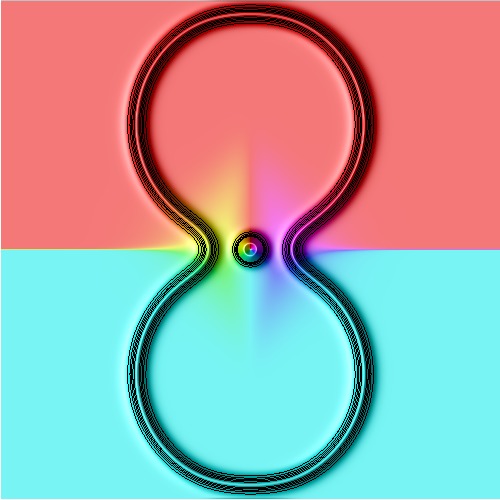} &
\includegraphics[scale=0.35,natwidth=1000,natheight=1000]{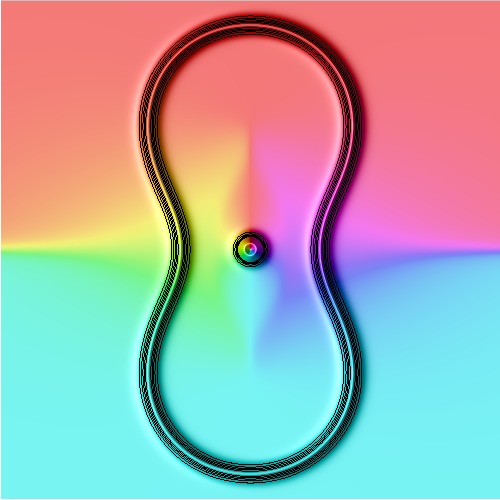} \\
\includegraphics[scale=0.35,natwidth=1000,natheight=1000]{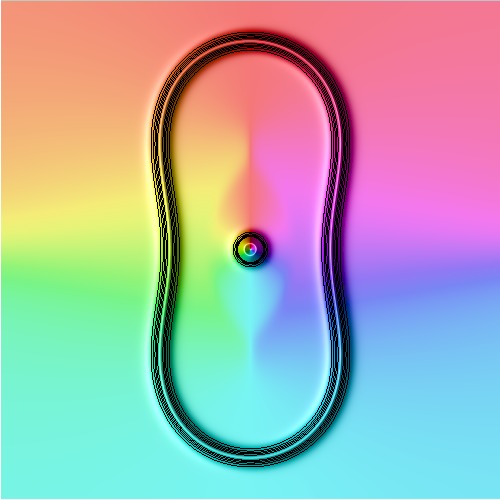} & \includegraphics[scale=0.35,natwidth=1000,natheight=1000]{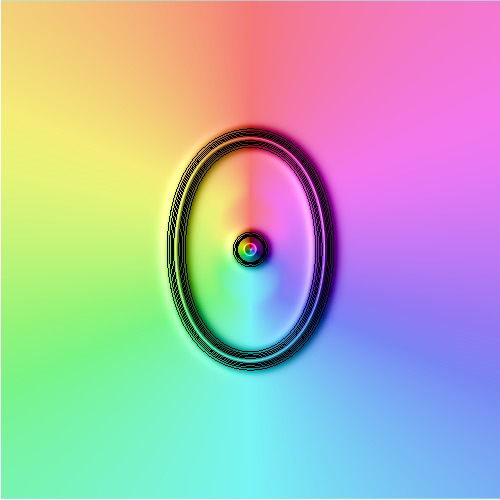} & \includegraphics[scale=0.35,natwidth=1000,natheight=1000]{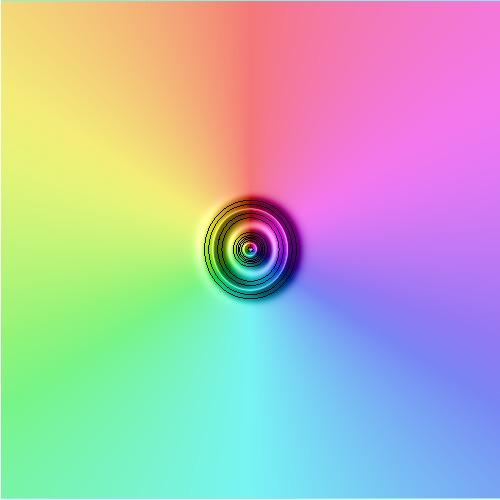} &
\includegraphics[scale=0.35,natwidth=1000,natheight=1000]{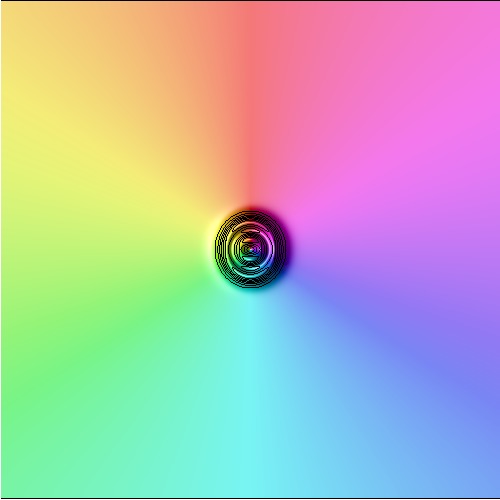}
\end{tabular}
\caption{Energy density plot of two domain wall bubbles meeting and forming a local winding and a baby Skyrmion. The wall has two points of fractional winding that cancel the interior baby Skyrmion. The fractional windings spread as the wall contracts ultimately annihilating with the interior baby Skyrmion. The initial conditions are highly constrained to produce the correct winding. The plot is coloured by the phase $\theta = \tan^{-1}\frac{\phi_2}{\phi_1}$.}
\label{2bubble}
\end{center}
\end{figure}

The numerical simulation of 2 domain wall loops interacting can be seen in figure \ref{2bubble}. The numerics here have a high damping term to ensure the baby Skyrmion is stable and to prevent the domain wall loops collapsing quickly. Note that while a large system is considered here, one may expect domain wall systems to be of an order much larger than the size of a single baby Skyrmion. The local charge density is created in the centre of the resulting domain loop, however the charge of the entire system remains zero. The domain wall then collapses in on the baby Skyrmion, ultimately annihilating.

\begin{figure}
\begin{center}
\begin{tabular}{c c c c c}
\includegraphics[scale=0.25,natwidth=1000,natheight=1000]{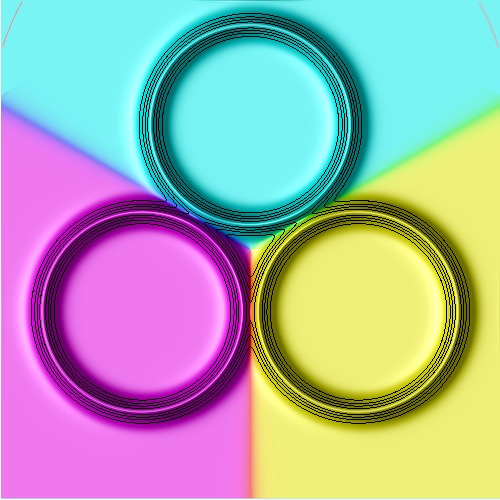} & \includegraphics[scale=0.25,natwidth=1000,natheight=1000]{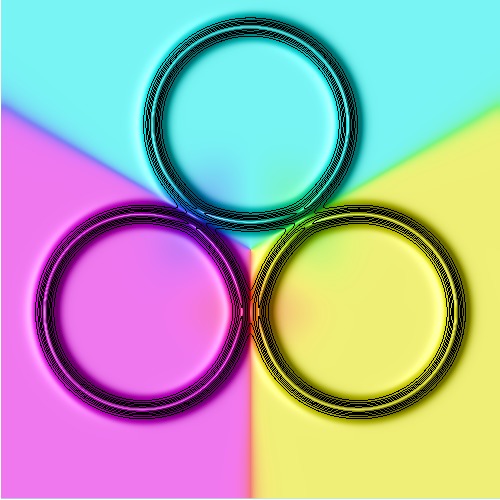} & \includegraphics[scale=0.25,natwidth=1000,natheight=1000]{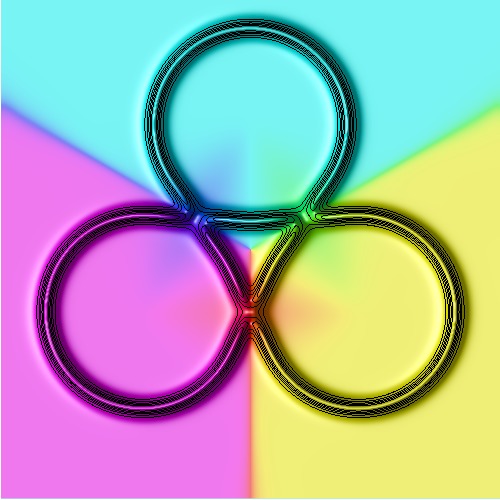} &
\includegraphics[scale=0.25,natwidth=1000,natheight=1000]{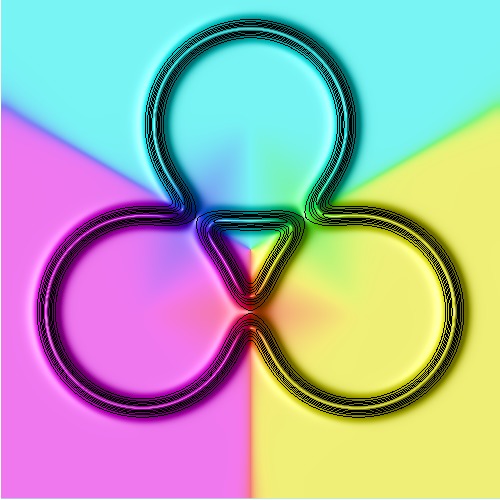} &
\includegraphics[scale=0.25,natwidth=1000,natheight=1000]{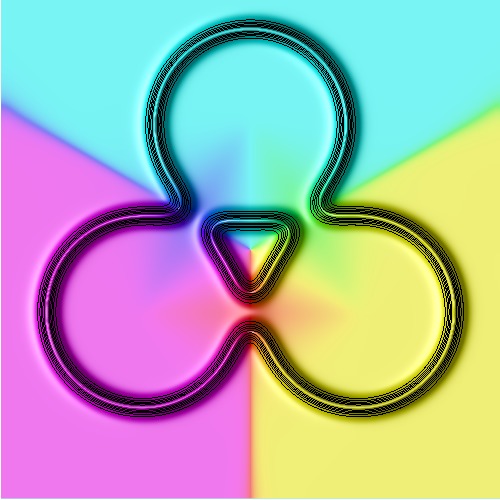} \\
\includegraphics[scale=0.25,natwidth=1000,natheight=1000]{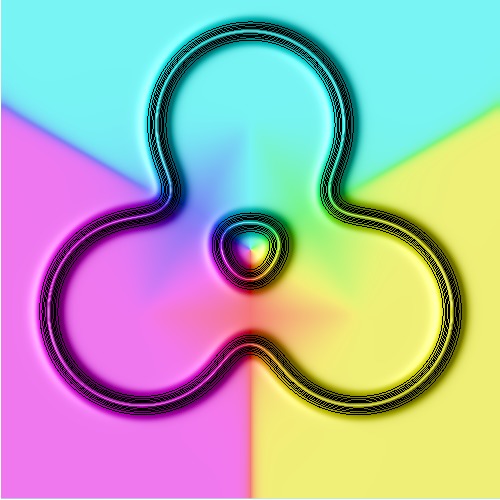} & \includegraphics[scale=0.25,natwidth=1000,natheight=1000]{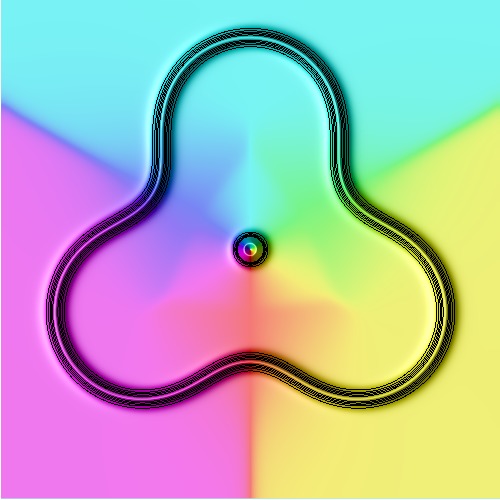} &
\includegraphics[scale=0.25,natwidth=1000,natheight=1000]{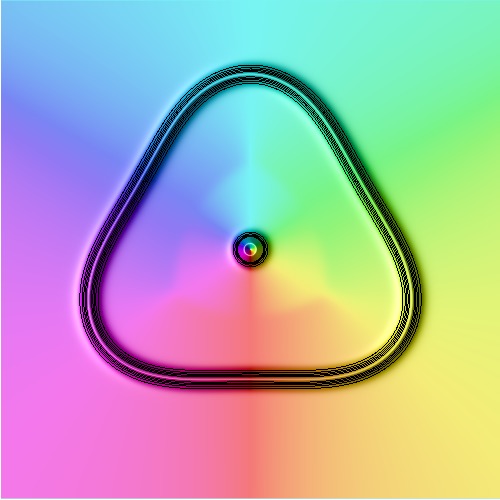} & 
\includegraphics[scale=0.25,natwidth=1000,natheight=1000]{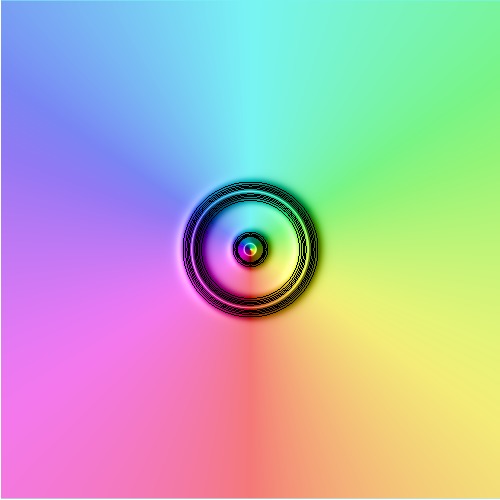} &
\includegraphics[scale=0.25,natwidth=1000,natheight=1000]{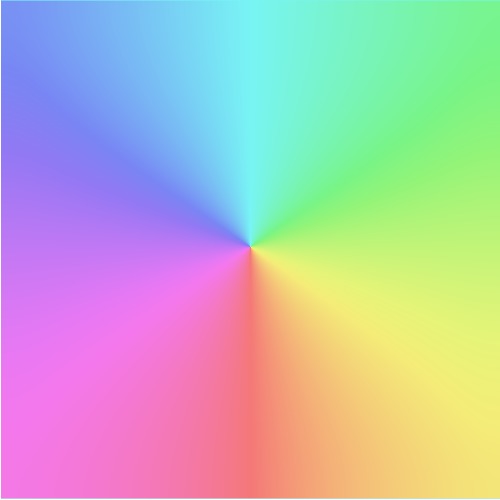} 
\end{tabular}
\caption{Energy density plot of three domain wall bubbles meeting and forming a local winding and a baby Skyrmion. The boundary then has three points of fractional winding that cancel the interior baby Skyrmion. The fractional windings spread as the wall contracts ultimately annihilating with the interior baby Skyrmion to the vacuum. The plot is coloured by the phase $\theta = \tan^{-1}\frac{\phi_2}{\phi_1}$.}
\label{3bubble}
\end{center}
\end{figure}

\begin{figure}
\begin{center}
\begin{tabular}{c c c}
\includegraphics[scale=0.35,natwidth=1000,natheight=1000]{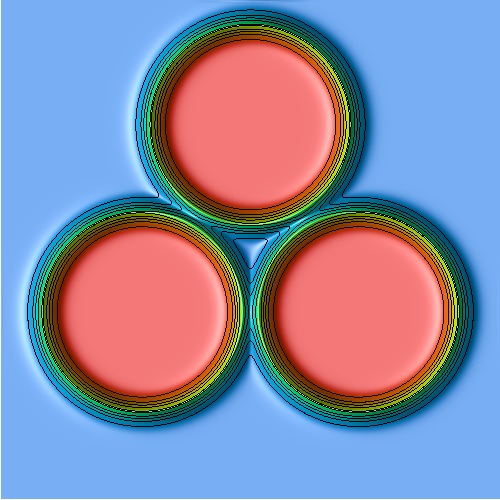} & \includegraphics[scale=0.35,natwidth=1000,natheight=1000]{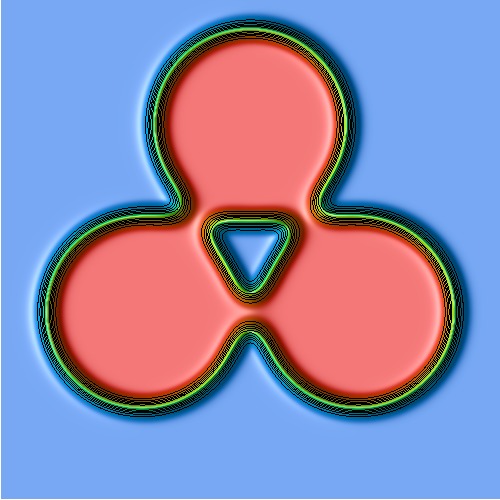} & \includegraphics[scale=0.35,natwidth=1000,natheight=1000]{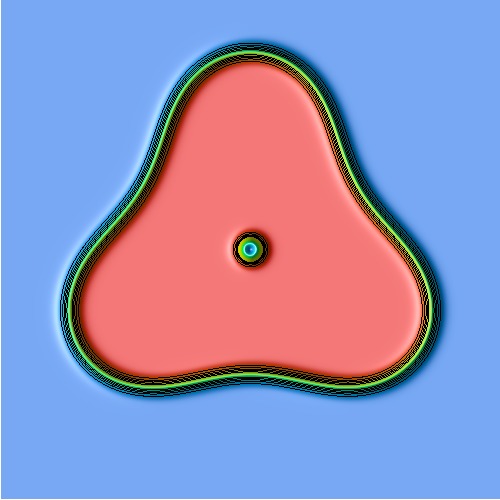}
\end{tabular}
\caption{Energy density plot of three domain wall bubbles meeting and forming a local winding and a baby Skyrmion. It is coloured by the $\phi_3$ value to show the vacua structure of the system at various constant time slices. The plots correspond with the simulation in figure \ref{3bubble}.}
\label{3bubble_phi3}
\end{center}
\end{figure}

A less constrained case is modelled with 3 domain wall bubbles meeting at various points in figure \ref{3bubble} (here quite symmetrically, though this is merely a product of minimising the size of the grid used). This time 3 bridges are formed, these meet and form a baby Skyrmion at the centre of the system. The bridges create a partial winding on the surrounding domain wall loop that spread out. Ultimately the domain wall loop shrinks and annihilates with the interior baby Skyrmion. The values for $\phi_3$ are also shown in figure \ref{3bubble_phi3} to demonstrate the vacua structure at various times of the simulation.

\begin{figure}
\begin{center}
\begin{tabular}{c c c c c}
\includegraphics[scale=0.25,natwidth=1000,natheight=1000]{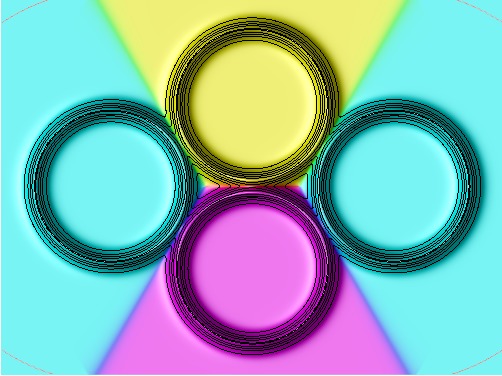} & \includegraphics[scale=0.25,natwidth=1000,natheight=1000]{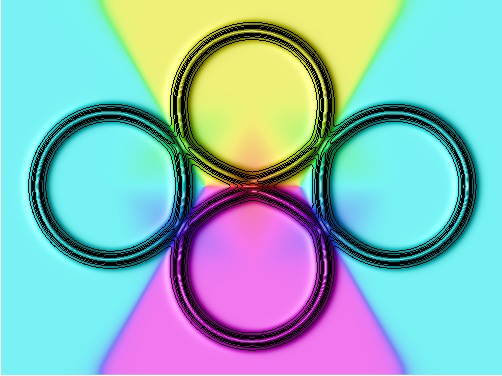} & \includegraphics[scale=0.25,natwidth=1000,natheight=1000]{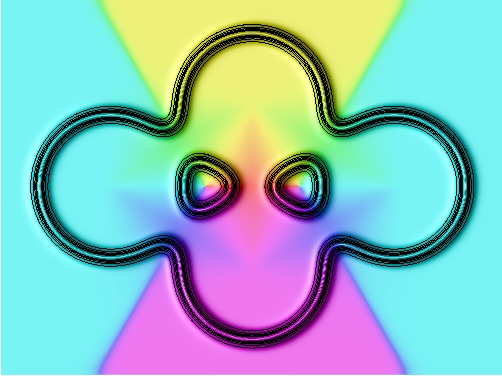} &
\includegraphics[scale=0.25,natwidth=1000,natheight=1000]{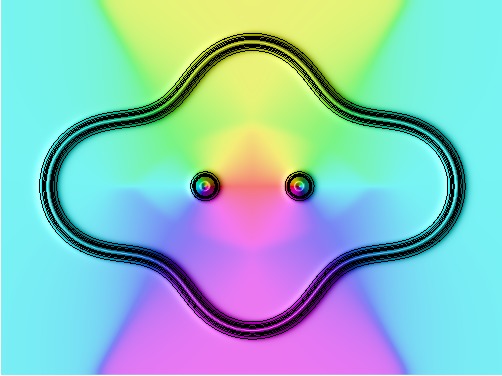} &
\includegraphics[scale=0.25,natwidth=1000,natheight=1000]{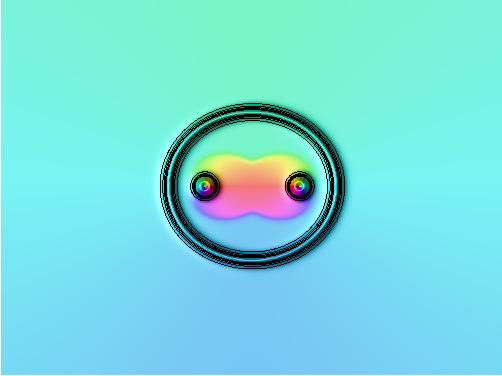} \\
\includegraphics[scale=0.25,natwidth=1000,natheight=1000]{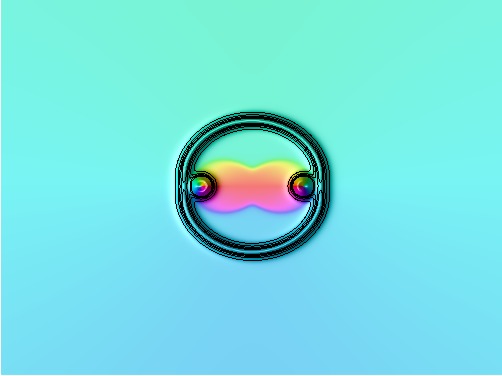} & \includegraphics[scale=0.25,natwidth=1000,natheight=1000]{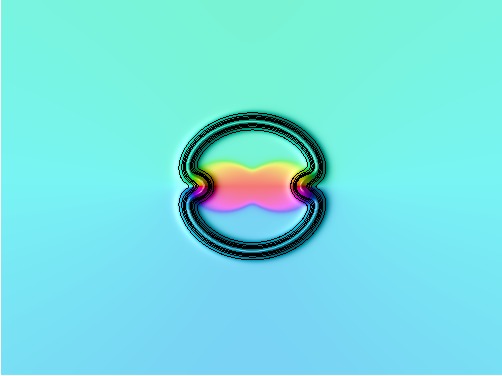} &
\includegraphics[scale=0.25,natwidth=1000,natheight=1000]{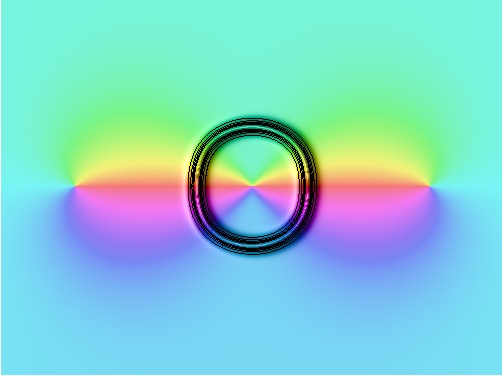} & 
\includegraphics[scale=0.25,natwidth=1000,natheight=1000]{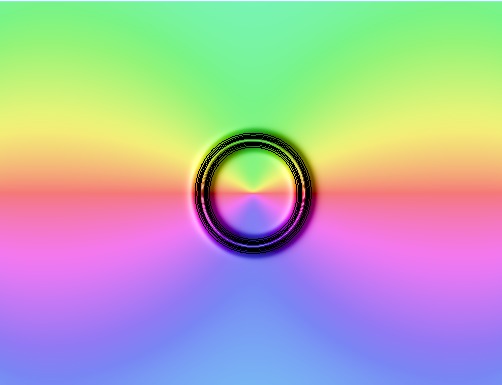} &
\includegraphics[scale=0.25,natwidth=1000,natheight=1000]{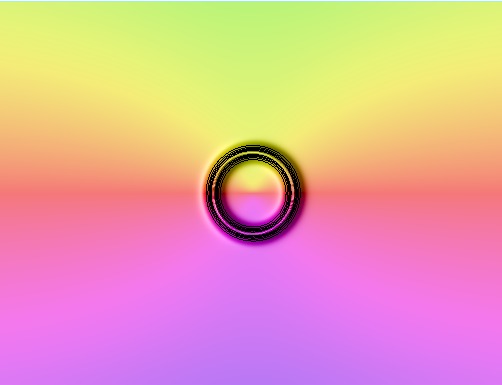} 
\end{tabular}
\caption{Energy density plot of four domain wall bubbles interacting to form a soliton and anti-soliton. The boundary has no resulting winding as the local charge of the soliton anti-soliton pair cancel. The two solitons are absorbed into the wall, with their winding then subsequently annihilating round the wall. The plot is coloured by the phase $\theta = \tan^{-1}\frac{\phi_2}{\phi_1}$.}
\label{4bubble}
\end{center}
\end{figure}

The final simulation, seen in figure \ref{4bubble} demonstrates 4 bubbles meeting to form a soliton and anti-soltion. The fractional windings annihilate around the surrounding domain wall loops. The soliton and anti-soliton are well separated hence don't annihilate. The domain wall boundary collapses in absorbing the solitons into the wall. The windings then annihilate around the domain wall boundary as it collapses.

These simulations represent what may happen at the meeting of two domain wall bubbles. It is also possible that bubbles may meet in several places forming chains of Skyrmion anti-Skyrmion pairs, as with the examples in the previous section. This means in a large system of domain walls one would find a complicated system of local charge distributions, within some walls which have sections of fractional winding, that effectively shield the exterior space from observing any change in topological charge.

It would be interesting to consider whether the interior system and interactions could be represented by what occurs on the boundary in some way. It would also be interesting to consider how the fractional winding sections interact with each other when traversing the domain wall.  

\section{(3+1) Skyrme Model}
The Lagrangian density for an $SU(2)$ valued Skyrme field $U(t,\boldsymbol{x})$ is given by,
\begin{equation}
\mathcal{L} = -\frac{1}{2} Tr\left(R_\mu R^\mu\right) + \frac{1}{16} Tr\left(\left[R_\mu,R_\nu\right]\left[R^\mu,R^\nu\right]\right) - m_\pi^2 Tr\left(2(1_2) - \left(U+U^\dagger\right)\right)
\label{lagden}
\end{equation}
where $R_i = \left(\partial_i U\right) U^\dagger$ is the right $su\left(2\right)$ valued current and $m_\pi$ is a constant mass parameter. The associated energy for a static Skyrme field $U(\boldsymbol{x})$ is,

\begin{equation}
 E = \frac{1}{12\pi^2} \int \left\{ -\frac{1}{2}Tr\left(R_i R^i\right) - \frac{1}{16}Tr\left(\left[R_i,R_j\right]\left[R^i,R^j\right]\right) + m^2_\pi Tr\left(2(1_2) - \left(U + U^\dagger\right)\right)  \right\} d^3 x
 \label{energy}
\end{equation}

Note that both of the above expressions have the parameters preceding the first two terms scaled out. We can reclaim a similar formulation as the $(2+1)$ model by writing everything in terms of the pion fields, using the $SU\left(2\right)$ nature of the field $U = \sigma + i\boldsymbol{\pi}\cdot\boldsymbol{\tau}$, where $\boldsymbol{\tau}$ is the triplet of Pauli matrices and $\boldsymbol{\pi} = \left(\pi_1, \pi_2, \pi_3\right)$ the triplet of pion fields. 

We have imposed a mass term to give us two vacua, noted as $U_\pm = \pm 1_2$. For finite energy we again require $\lim_{\left|\boldsymbol{x}\right|\rightarrow \infty}U = U_\pm$. This gives us the map $U:\mathbb{R}^3\cup \left\{\infty\right\} = S^3 \rightarrow S^3$, and hence a topological charge as an element of the 3rd homotopy group, equivalent to an integer $B \in \pi_3\left(S^3\right) = \mathbb{Z}$,

\begin{equation} 
B = -\frac{1}{24\pi^2}\int \epsilon_{ijk}Tr\left(R_i R_j R_k\right) d^3 x.
\end{equation}

It is proposed that Skyrmions can be formed in a similar manner to the planar model, from the collision of two domain walls. Here again domain walls are energy configurations arising from the interpolation between the two vacuum sates, namely $U_\pm$.

\section{Skyrmion Formation Examples}
\begin{center}
\begin{figure}
\begin{center}
\begin{tabular}{c c c c}
\begin{overpic}[trim=50 20 0 20,clip,scale=0.5]{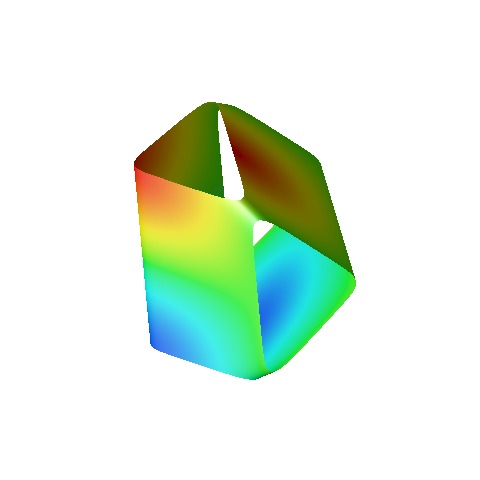}
 \put (34,0) {\large$\displaystyle\pi_1$}
\end{overpic} & \begin{overpic}[trim=50 20 0 20,clip,scale=0.5]{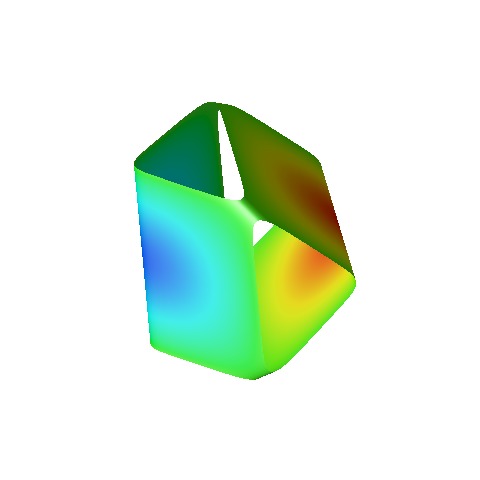}
 \put (34,0) {\large$\displaystyle\pi_2$}
\end{overpic} &
\begin{overpic}[trim=50 20 0 20,clip,scale=0.5]{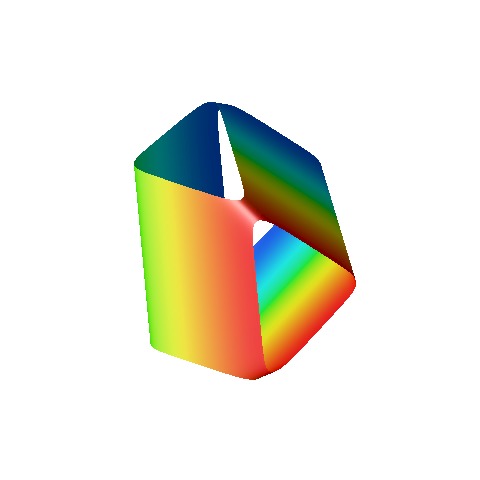}
 \put (34,0) {\large$\displaystyle\pi_3$}
\end{overpic} &
\includegraphics[trim=0 40 0 40,clip,scale = 0.3]{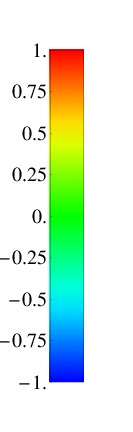}
\end{tabular}
\end{center}
\caption{Initial conditions of two domain walls meeting, used to form a single soliton for the full $SU(2)$ Skyrme model, isosurface of $\sigma = 0$ with colours based on the value of $\pi_1, \pi_2, \pi_3$ respectively. The final panel shows the colourbar for the values each colour represents for the respective pion field. }
\label{Initialconditions}
\end{figure}
\end{center}

\begin{center}
\begin{figure}
\begin{center}
\begin{tabular}{c c c c c}
\includegraphics[trim=50 20 50 20,clip,scale=0.4,natwidth=1000,natheight=1000]{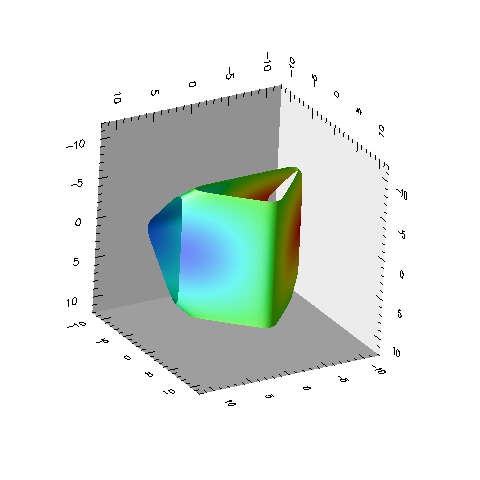} & \includegraphics[trim=50 20 50 20,clip,scale=0.4,natwidth=1000,natheight=1000]{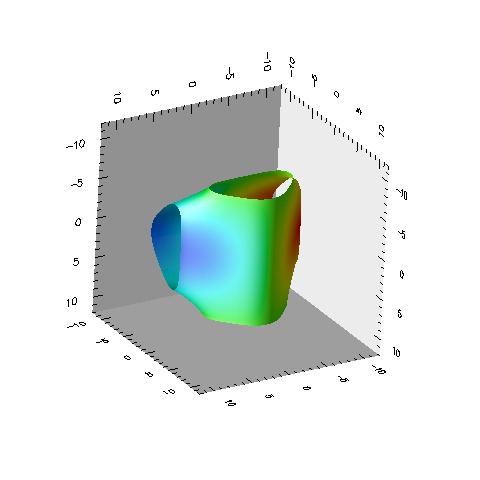} & \includegraphics[trim=50 20 50 20,clip,scale=0.4,natwidth=1000,natheight=1000]{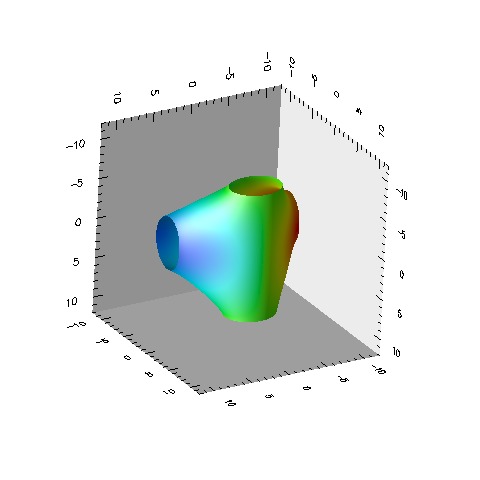} &
\includegraphics[trim=50 20 50 20,clip,scale=0.4,natwidth=1000,natheight=1000]{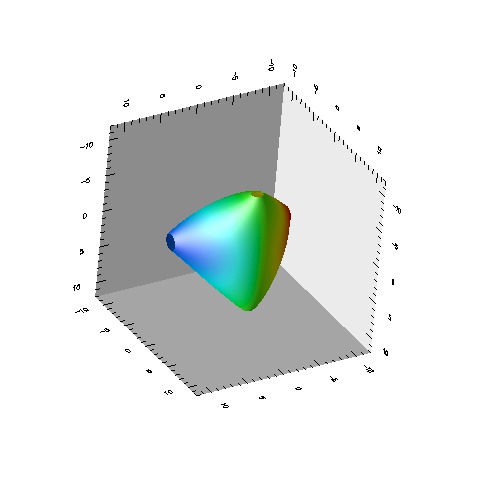} &
\includegraphics[trim=50 20 50 20,clip,scale=0.4,natwidth=1000,natheight=1000]{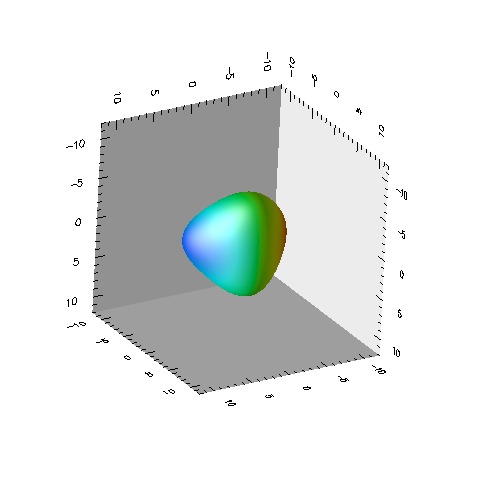}\\
$t = 0$ & $t = 8.32$ & $t = 41.6$ & $t = 56.16$ & $t = 60.32$ \\
\end{tabular}
\begin{tabular}{c c c c}
\includegraphics[trim=50 20 50 20,clip,scale=0.4,natwidth=1000,natheight=1000]{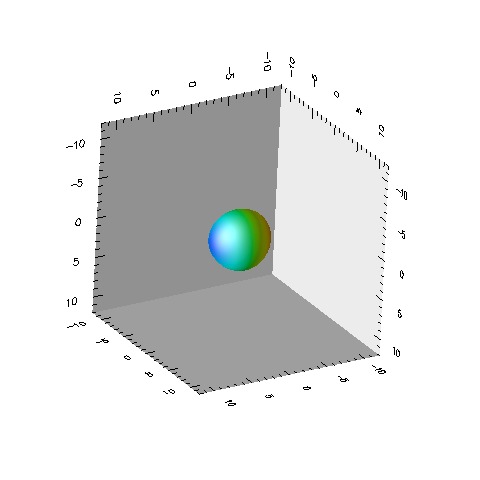} & \includegraphics[trim=50 20 50 20,clip,scale=0.4,natwidth=1000,natheight=1000]{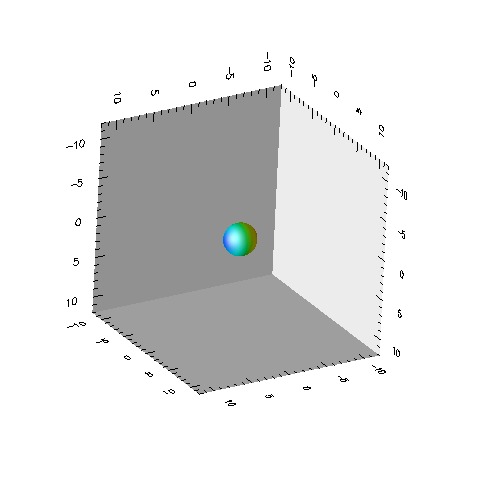} &
\includegraphics[trim=50 20 50 20,clip,scale=0.4,natwidth=1000,natheight=1000]{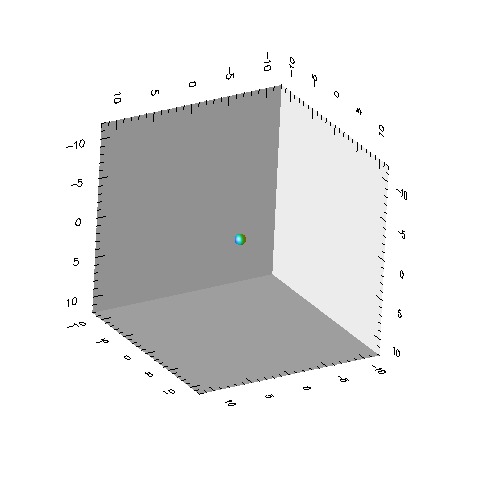} &
\includegraphics[trim=50 20 50 20,clip,scale=0.4,natwidth=1000,natheight=1000]{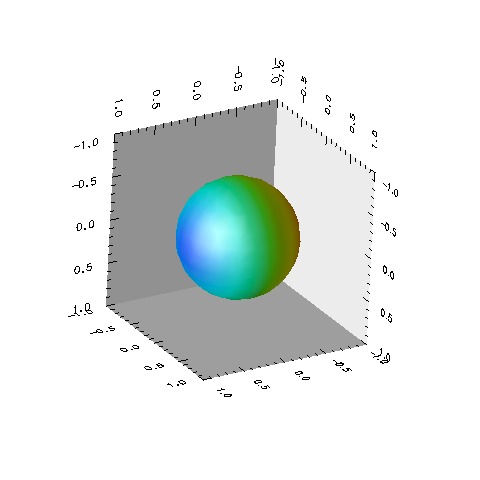}\\
$t = 95.68$ & $t = 116.48$ & $t = 137.28$ & $t = 200$
\end{tabular}
\caption{Simulation of two domain walls meeting to form a single soliton. The initial conditions (given in figure \ref{Initialconditions}) are highly constrained. The plot is an isosurface of $\sigma = 0$ with colours based on the value of $\pi_1$ (colours match the colour bar in figure \ref{Initialconditions}). The final panel is the resulting stable Skyrmion blown up so it is visible, the configuration matches the previous panel.}
\label{3dformation}
\end{center}
\end{figure}
\end{center}

\begin{figure}[h]
\begin{center}
\begin{tabular}{c c c}
\includegraphics[trim=20 0 20 0,clip,scale=0.4,natwidth=1000,natheight=1000]{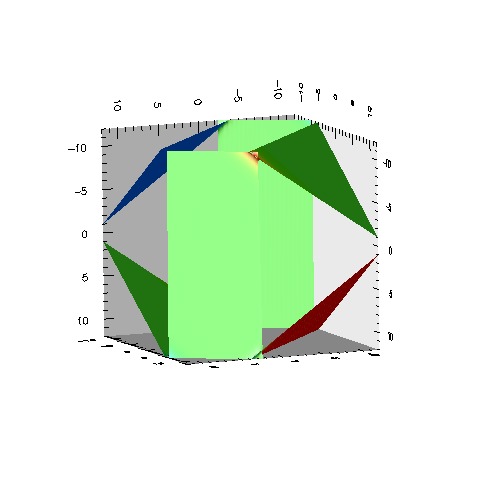} & \includegraphics[trim=20 0 20 0,clip,scale=0.4,natwidth=1000,natheight=1000]{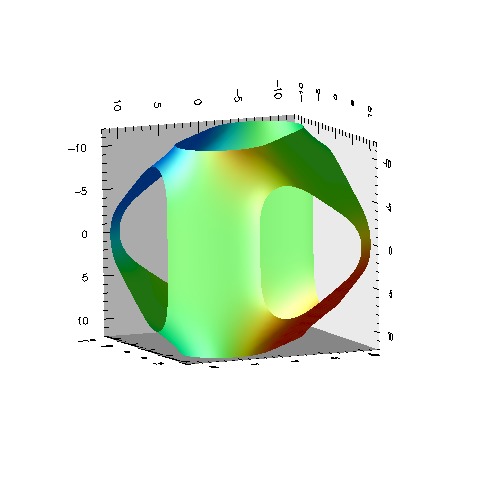} & \includegraphics[trim=20 0 20 0,clip,scale=0.4,natwidth=1000,natheight=1000]{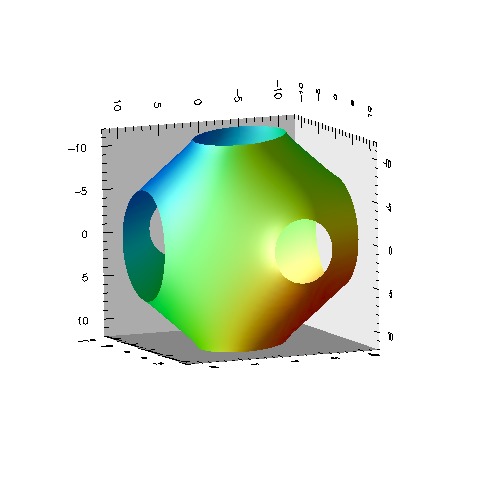}\\
$t = 0$ & $t = 12.8$ & $t = 57.6$ \\
\end{tabular}
\begin{tabular}{c c c c}
\includegraphics[trim=20 0 20 0,clip,scale=0.4,natwidth=1000,natheight=1000]{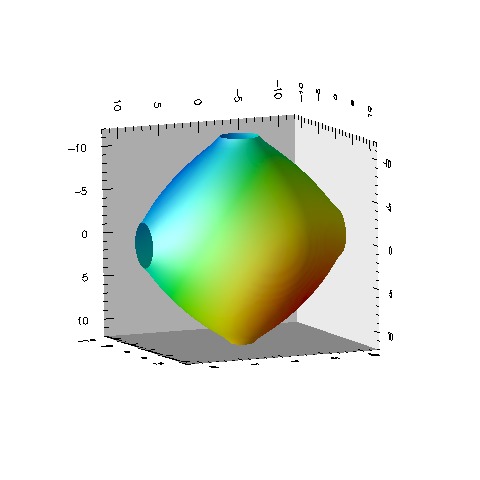} & \includegraphics[trim=20 0 20 0,clip,scale=0.4,natwidth=1000,natheight=1000]{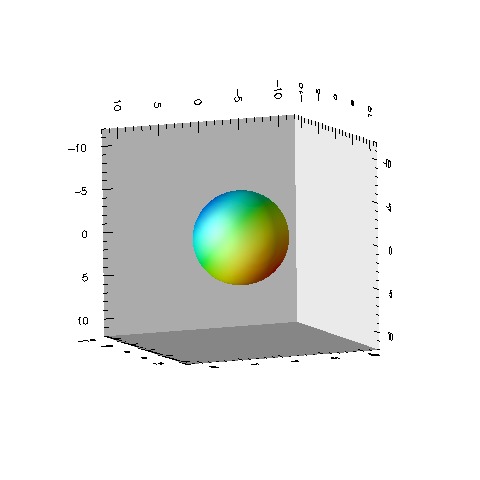} &
\includegraphics[trim=20 0 20 0,clip,scale=0.4,natwidth=1000,natheight=1000]{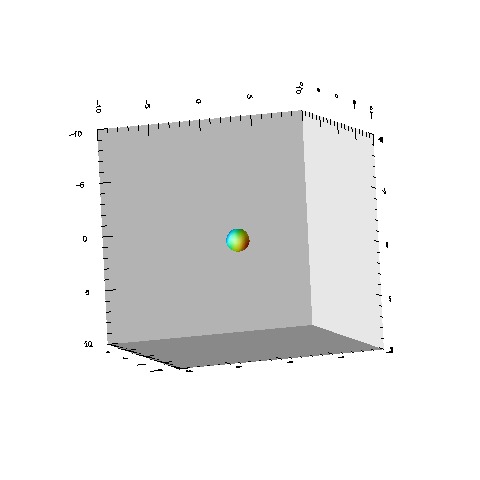}&
\includegraphics[trim=50 0 50 0,clip,scale=0.4,natwidth=1000,natheight=1000]{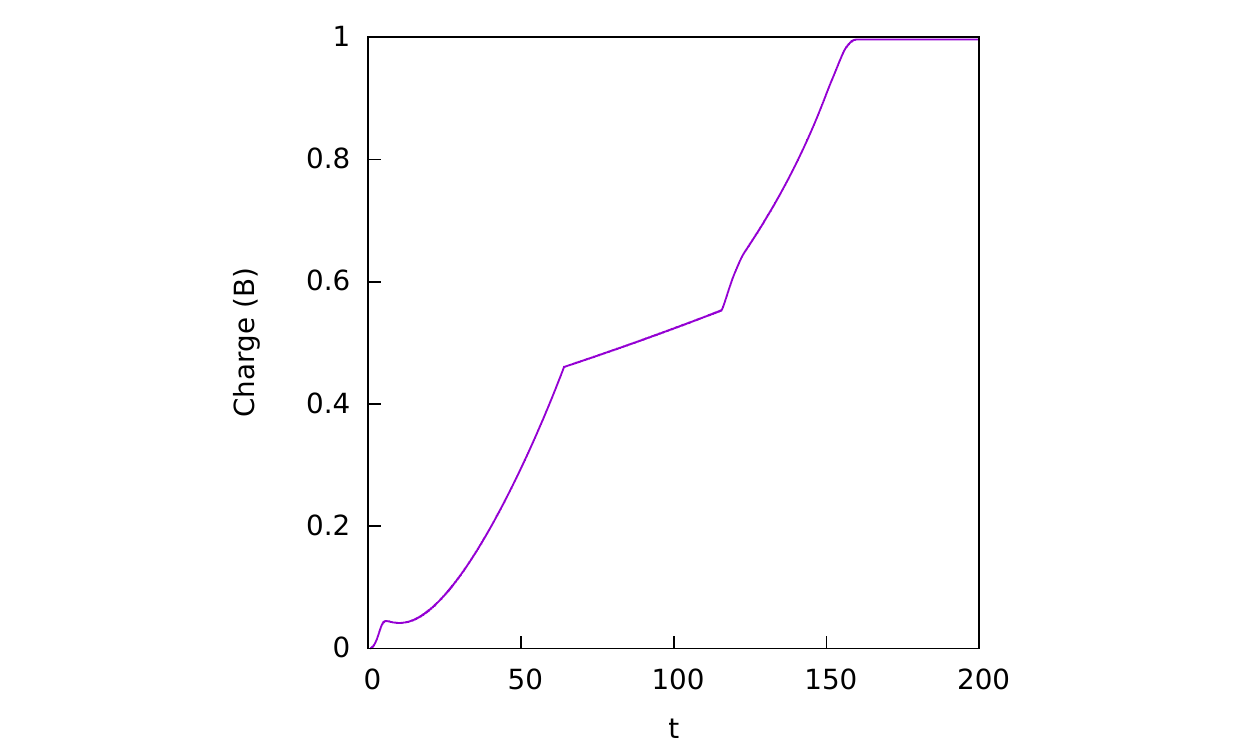} \\
$t = 153.6$ & $t = 204.8$ & $t = 227.2$
\end{tabular}
\caption{Isosurface plot for $\sigma = 0$ demonstrating 6 domain walls forming a single Skyrmion, coloured by the value of $\pi_1$. The topological charge is given in the final panel.}
\label{3dwalls}
\end{center}
\end{figure}

Simulations of the nonlinear time-dependent PDE that follows from the variation of \ref{lagden} were performed using a fourth order Runge-Kutta method on a grid of 101x101x101 grid points. We used Neumann boundary conditions (the spatial derivative normal to the boundary vanishes), which again allows the domain walls to move unhindered. We first simulate the proposed formation method of two incident domain walls. The initial conditions have to be more constrained than in the planar case and can be seen in figure \ref{Initialconditions}. The addition of an extra field as well as an additional dimension, makes producing the correct winding quite challenging. The formation process can be observed in figure \ref{3dformation}.
We now present a similar solution as the planar case, with multiple incident domain walls. Due to the additional difficulties in producing the correct winding, we have used 6 domain walls, to produce the required affect, which can be seen in figure \ref{3dwalls}. This should be attainable using a fewer number of domain walls however the simulations are challenging to set up (partly this is due to the field not being able to change in the corner of the simulation with our chosen boundary conditions). 
Finally, a $(3+1)$ domain wall system is extremely difficult to simulate. However the results should follow a similar form to the results presented for the $(2+1)$ dimensional system. The main difference is the increased difficulty in forming the correct conditions for the correct winding of all 3 fields. Though the increased computing power needed due to the additional spatial dimension is also somewhat restrictive.  

\section{Conclusions}
We have demonstrated several situations in which Skyrmion solutions can be produced by domain wall interactions in both the $(2+1)$-dimensional baby Skyrme model and the $(3+1)$-dimensional $SU(2)$ Skyrme model. We also demonstrated that using more than 2 domain walls, decreases the required constraints on the system for formation to occur. It is possible that these techniques could be utilised in condensed matter systems to produce Skyrmions. We have also modelled the interactions of domain wall networks, demonstrating how Skyrmions can be formed within these. It was shown that for the topoloigcal charge to remain conserved, a counteracting winding was formed along the boundary of the system.

This paper has raised a few interesting questions that have gone unanswered here. Firstly, how feasible would this method be for forming Skyrmions in a condensed matter system at a bifurcation point (Y-junction). Also, could a condensed matter system be used to give the $D_N$ symmetry to the incident domain walls, to increase the probability of formation to occur. Secondly, does the counteracting winding on the boundary of a domain wall system, allow any information regarding the interior winding to be attained. To understand this we are likely to need to understand the nature of interactions of the bridges, or fractional winding segments that propagate around the boundary. Finally it would be interesting to be able to make some statistical predictions on the formations of Skyrmions in a large domain wall network. This may also be able to be related to the excitation of a vacuum state of a system, to see if Skyrmions could be formed this way, in a non-perturbative manner.

\section{Acknowledgements}
I would like to thank EPSRC for funding me during my PhD. I would also like to thank my supervisor Paul Sutcliffe for useful discussions.

\bibliographystyle{prsty}
\bibliography{Formation}

\begin{thebibliography}{10}

\bibitem{Skyrme:1961vq}
T. Skyrme, Proc.Roy.Soc.Lond. {\bf A260},  127  (1961).

\bibitem{bible}
N. Manton and P. Sutcliffe, {\em Topological Solitons} (Cambridge University
  Press, Cambridge, 2004).

\bibitem{Battye:2001qn}
R.~A. Battye and P.~M. Sutcliffe, Rev.Math.Phys. {\bf 14},  29  (2002).

\bibitem{Piette:1994ug}
B.~M. A.~G. Piette, B.~J. Schroers, and W. Zakrzewski, Z.Phys. {\bf C65},  165
  (1995).

\bibitem{Sondhi:1993zz}
S. Sondhi, A. Karlhede, S. Kivelson, and E. Rezayi, Phys.Rev. {\bf B47},  16419
   (1993).

\bibitem{Yu}
X.~Z. Yu {\it et~al.}, Nature {\bf 465},  901  (2010).

\bibitem{sampaio2013nucleation}
J. Sampaio {\it et~al.}, Nature nanotechnology {\bf 8},  839  (2013).

\bibitem{iwasaki2013current}
J. Iwasaki, M. Mochizuki, and N. Nagaosa, Nature nanotechnology {\bf 8},  742
  (2013).

\bibitem{Nitta:2012xq}
M. Nitta, Phys.Rev. {\bf D86},  125004  (2012).

\bibitem{Kobayashi:2013ju}
M. Kobayashi and M. Nitta, Phys.Rev. {\bf D87},  085003  (2013).

\bibitem{PhysRevA.85.053639}
M. Nitta, K. Kasamatsu, M. Tsubota, and H. Takeuchi, Phys. Rev. A {\bf 85},
  053639  (2012).

\bibitem{Jaykka:2011ic}
J. Jaykka, M. Speight, and P. Sutcliffe, Proc.Roy.Soc.Lond. {\bf A468},  1085
  (2012).

\bibitem{Winyard:2013ada}
P. Jennings and T. Winyard, JHEP {\bf 1401},  122  (2014).

\end{thebibliography}
\end{document}